\documentclass[conference,letterpaper]{IEEEtran}

\usepackage[utf8]{inputenc} 
\usepackage[T1]{fontenc}
\usepackage{url}
\usepackage{ifthen}
\usepackage{cite}
\usepackage[cmex10]{amsmath} 
\usepackage[pdftex]{graphicx}
\usepackage{amsmath}
\usepackage{amssymb}
\usepackage{amsfonts}
\usepackage{latexsym}
\usepackage{bm}
\usepackage{float}
\usepackage{nidanfloat}
\usepackage{wrapfig}
\usepackage{moreverb}
\usepackage{ascmac}
\usepackage{mathtools}

\newtheorem{dfn}{Definition}
\newtheorem{thm}{Theorem}
\newtheorem{rem}{Remark}
\newtheorem{lem}{Lemma}

\begin{document}
\title{Unified Expression of Utility-Privacy Trade-off\\ in Privacy-Constrained Source Coding} 


\author{%
  \IEEEauthorblockN{Naruki Shinohara and Hideki Yagi}
  \IEEEauthorblockA{Dept. of Computer and Networking Engineering\\
                    The University of Electro-Communications, Tokyo, Japan\\
                    Email: s1710291@mail.uec.jp, h.yagi@uec.ac.jp}
}


\maketitle

\begin{abstract}
Privacy-constrained source coding problems have become increasingly important recently, and the utility-privacy trade-off has been investigated for various systems. As pioneering work, Yamamoto (1983) found theoretical limits of the coding rate, privacy and utility in two cases; (i) both public and private information is encoded and (ii) only public information is encoded. However, the theoretical limit has not been characterized in a more general case; (iii) encoded messages consist of public information and a part of private information. Then in this paper, we characterize the trade-off relation in case (iii) using rate-distortion theory. The obtained expression of the achievable region is a ``unified expression'' because it includes the ones in cases (i) and (ii) as special cases. The numerical results also demonstrate that neither case (i) nor (ii) are the best cases, and it is important to select the encoded information adequately in case (iii).  
\end{abstract}

\section{Introduction}\label{sec:1}

According to the development of information-technology, the amount of data processed in devices increases rapidly. At the same time, the risk of accidental or intentional leakage of private information increases quickly, too. Therefore, Yamamoto \cite{a} claimed the necessity of converting databases to maximize utility of data while maintaining privacy.
He revealed the optimal relationships (theoretical limits) among coding rate, utility, and privacy in two cases; (i) public information that can be open to public and private information that should be protected from a third party are encoded, and (ii) only public information is encoded. Numerical results for binary source indicate that the optimal privacy-leakage rate in case (i) is smaller than the one in case (ii) while for the optimal privacy-leakage rates, the achievable coding rate in case (ii) is smaller than the one in case (i). From this fact, considering another case; (iii) public information and a part of private information is encoded, it is expected that the optimal privacy in cases (i) and (iii) is securer than the one in case (ii) whereas for the optimal privacy, the achievable coding rates in cases (i) and (iii) are larger than the one in case (ii). Subsequently, Sankar et al. \cite{b} analyzed the coding system in which the encoder can observe side information and likewise characterized the theoretical limits among the coding rate, utility, and privacy. The problem related to utility-privacy trade-off has extensively studied in \cite{g}--\!\!\cite{k}.

As a general setting, encoded messages may consist of public information and a part of private information (case (iii)). From the practical aspect, the analysis of case (iii) is beneficial because some information givers may hope to achieve high security against not only a third party but also aggregators (encoder). However, it is still unclear how the region is characterized in case (iii). Since case (iii) includes both cases (i) and (ii) as special instances, it is of importance to provide complete characterization of the fundamental limits.

In this paper, we analyze the trade-off relationship among the coding rate, utility, and privacy in case (iii) based on rate-distortion theory. For that purpose, we take advantage of one concept about the encoded information, that is, the encoded set (see Sect.\hspace{1mm}I\hspace{-.1em}I). Our characterization of the achievable region gives a ``unified expression'' because it includes the characterizations given in \cite{a} in cases (i) and (ii) as special cases. In addition, the unified expression of the theoretical limit enables us to describe the achievable regions of several cases concisely. Through the numerical results, we clarify the significance of case (iii).

From the results of this paper, it is confirmed that finding the theoretical limit in case (iii) is formulated as a convex programming problem. It is proved that when the Arimoto-Blahut algorithm \cite{c}, \cite{d} is applied to convex programming problems, the alternating procedure converges. Therefore, it may be effective to calculate the theoretical limit. 

In the proof, the converse part is based on the standard argument in information theory \cite{m}. The direct part is a bit more challenging. We combine the type method (Csisz\'ar-K\"{o}rner method) \cite{f} with the random coding method (Wyner-Ziv method) \cite{m}. Since the encoded information may be different from either public information or all information, the evaluation of privacy is done by introducing two typical sets conditioned by the encoded information sequence. One set represents the set of the all information sequences which include the encoded information as subsequences (\eqref{eq:4.21}). The other set represents the set for which an additional constraint is further imposed on the first typical set (\eqref{eq:4.22}). Then, by the strong typicality, these typical sets lead to a tight lower bound on privacy.

\section{Notation and System Model}

\subsection{Source}

Database $d$ is described by a $K \times n$ matrix whose rows represent $K$ attributes and columns represent $n$ entries of data. Let $\mathcal{K} = \{1, 2, \ldots , K\}$ be the set of indexes of $K$ attributes. The random variable for the $l$th attribute is denoted by $X_{l}$, which takes a value in a finite alphabet $\mathcal{X}_{l}$. For any subset $\mathcal{B} \subseteq \mathcal{K}$, the tuple of random variables $(X_{l})_{l \in \mathcal{B}}$ is abbreviated as $X_{\mathcal{B}}$. Similarly, the Cartesian product of alphabets $\prod_{l \in \mathcal{B}} \mathcal{X}_{l}$ is abbreviated as $\mathcal{X}_{\mathcal{B}}$.

The $K$ attributes can be divided into two groups; one may be open to public and the other should be kept secret from a third party. Then, the set $\mathcal{K}$ is divided into disjoint sets $\mathcal{R}$ and $\mathcal{H}$. That is,
\begin{align}
\mathcal{K}=\mathcal{R}\cup\mathcal{H},~~~~\mathcal{R}\cap\mathcal{H}=\emptyset,~~~~\mathcal{X}_{\mathcal{K}}=\mathcal{X}_{\mathcal{R}}\times\mathcal{X}_{\mathcal{H}}, \label{eq:2.1}
\end{align}
where $\mathcal{X}_{\mathcal{R}}$ is the set of values in which public (revealed) source symbols $X_{\mathcal{R}}$ take and $\mathcal{X}_{\mathcal{H}}$ is the set of values in which private (hidden) source symbols $X_{\mathcal{H}}$ take.

We assume that the source sequence $X_{\mathcal{K}}^{n} = (X_{\mathcal{K}, 1}, X_{\mathcal{K}, 2}, \ldots, X_{\mathcal{K}, n})$ is generated from a stationary and memoryless source $p_{X_{\mathcal{K}}}$.
That is,
\begin{align}
p_{X_{\mathcal{K}}^{n}}(x_{\cal{K}}^{n}) =  \Pr\{X_{\cal{K}}^{n} = x_{\cal{K}}^{n}\} = \prod_{i=1}^{n}p_{X_{\mathcal{K}}}(x_{\mathcal{K}, i}),\label{eq:2.6}
\end{align}
where $x_{\mathcal{K}}^{n} = (x_{\mathcal{K}, 1}, \ldots, x_{\mathcal{K}, n}) \in \mathcal{X}_{\mathcal{K}}^{n}$. Taking the partition of attributes in (\ref{eq:2.1}) into account, the source sequence $X_{\mathcal{K}}^{n}$ is described as
\begin{align}
X_{\mathcal{K}}^{n}=(X_{\mathcal{R}}^{n}, X_{\mathcal{H}}^{n}),\label{eq:2.2}
\end{align}
where
\begin{align}
&X_{\mathcal{R}}^{n}=(X_{\mathcal{R}, 1}, X_{\mathcal{R}, 2}, \ldots, X_{\mathcal{R}, n})\in\mathcal{X}_{\mathcal{R}}^{n},\label{eq:2.3}\\
&X_{\mathcal{H}}^{n}=(X_{\mathcal{H}, 1}, X_{\mathcal{H}, 2}, \ldots, X_{\mathcal{H}, n})\in\mathcal{X}_{\mathcal{H}}^{n}\label{eq:2.4}
\end{align}
are referred to as the revealed source sequence and the hidden source sequence, respectively. In the addressed coding system, the revealed symbols and a part of the hidden symbols are input to the encoder, and thus the encoded alphabet $\mathcal{E}$ satisfies $\mathcal{R} \subseteq \mathcal{E} \subseteq \mathcal{K}$. Similar to \eqref{eq:2.2}, $X_{\mathcal{K}}^{n}$ is sometimes described as
\begin{align}
X_{\mathcal{K}}^{n}=(X_{\mathcal{E}}^{n}, X_{\mathcal{E}^{\mathrm{c}}}^{n}),\label{eq:2.5}
\end{align}
where $X_{\mathcal{E}}^{n}$ is the source sequence observed by the encoder and $\mathcal{E}^{\mathrm{c}} = \mathcal{K} \setminus \mathcal{E}$.

\subsection{Encoder and Decoder}

\begin{figure}[H]
\begin{center}
\includegraphics[scale=0.48]{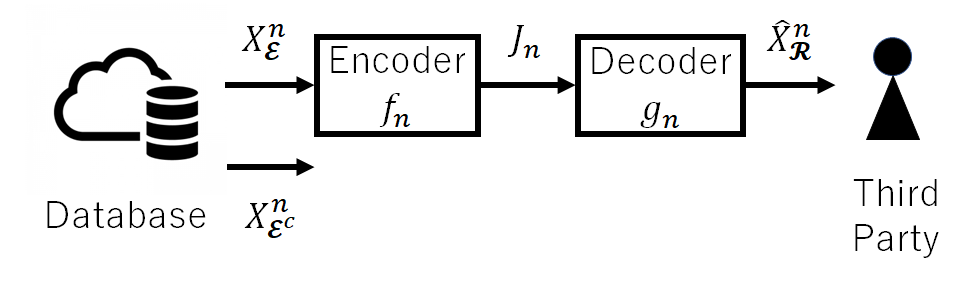}
\caption{Privacy-constrained coding system}
\label{Fig:1}
\end{center}
\end{figure}
The coding system consists of the encoder $f_{n}$ and the decoder $g_{n}$ as in Fig.\ref{Fig:1}. When the source sequence $X_{\mathcal{K}}^{n}=(X_{\mathcal{E}}^{n}, X_{\mathcal{E}^{\mathrm{c}}}^{n})$ is generated from the stationary and memoryless source $p_{X_{\mathcal{K}}}$, codeword $J_{n} = f_{n}(X_{\mathcal{E}}^{n})$ is generated by the encoder
\begin{align}
f_{n}\!:~\mathcal{X}_{\mathcal{E}}^{n}\rightarrow\{1, {2}, \ldots, M_{n}\}\label{eq:2.7}
\end{align}
and reproduced sequence $\hat{X}_{\mathcal{R}}^{n} = g_{n}(J_{n})$ is produced by the decoder
\begin{align}
g_{n}\!:~\{1, {2}, \ldots, M_{n}\}\rightarrow\hat{\mathcal{X}}_{\mathcal{R}}^{n},\label{eq:2.8}
\end{align}
where $M_{n}$ denotes the number of codewords.

\subsection{Measure of Coding Rate, Utility, and Privacy}

In this subsection, we mention the measure of the coding rate, utility, and privacy given in \cite{b}. Hereafter, let a pair of the encoder and decoder $(f_{n}, g_{n})$ be fixed.

For given $M_{n}$, the coding rate is defined as
\begin{align}
r_{n}\coloneqq\frac{1}{n}\log{M_{n}}.\label{eq:2.9}
\end{align}

Let $d:~\mathcal{X}_{\mathcal{R}}\times\hat{{\mathcal{X}}}_{\mathcal{R}}\rightarrow[0, \infty)$ be a distortion function between $x_{\mathcal{R}} \in \mathcal{X}_{\mathcal{R}}$ and $\hat{x}_{\mathcal{R}} \in \hat{\mathcal{X}}_{\mathcal{R}}$. The distortion between sequences $x_{\mathcal{R}}^{n} \in \mathcal{X}_{\mathcal{R}}^{n}$ and $\hat{x}_{\mathcal{R}}^{n} \in \hat{\mathcal{X}}_{\mathcal{R}}^{n}$ is defined as
\begin{align}
d(x_{\mathcal{R}}^{n}, \hat{x}_{\mathcal{R}}^{n})&\coloneqq\sum_{i=1}^{n}d(x_{\mathcal{R},i}, \hat{x}_{\mathcal{R},i}). \label{eq:2.0}
\end{align}
Then, the measure of utility is defined as
\begin{align}
u_{n}&\coloneqq\mathbb{E}\left[\frac{1}{n}d(X_{\mathcal{R}}^{n}, \hat{X}_{\mathcal{R}}^{n})\right], \label{eq:2.10}
\end{align}
where $\mathbb{E}$ represents the expectation by the joint distribution of $(X_{\mathcal{R}}^{n}, \hat{X}_{\mathcal{R}}^{n})$.

In this system, privacy of the hidden source sequence $X_{\mathcal{H}}^{n}$ should be protected when the codeword $J_{n}$ is observed by a third party. The measure of privacy is defined as
\begin{align}
e_{n}\coloneqq\frac{1}{n}H(X_{\mathcal{H}}^{n}|J_{n}), \label{eq:2.12}
\end{align}
where $H(X_{\mathcal{H}}^{n} | J_{n})$ is the conditional entropy of $X_{\mathcal{H}}^{n}$ given $J_{n}$.

\section{Problem Formulation and Main Theorem}
\label{sec:3}

\subsection{Achievable Region}

In this subsection, we define the achievable region investigated in this study. Moreover, we review known results given by related studies.
\begin{dfn}
A tuple $(R, D, E)$ is said to be {\bf achievable} if for any given $\epsilon > 0$, there exists a sequence of codes $(f_{n}, g_{n})$ satisfying
\begin{align}
r_{n}&\le R+\epsilon, \label{eq:3.1}\\
u_{n}&\le D+\epsilon, \label{eq:3.2}\\
e_{n}&\ge E-\epsilon \label{eq:3.3}
\end{align}
for all sufficiently large $n$.
\hspace{\fill}$\Box$\label{dfn:1}
\end{dfn}

\begin{rem}
From Definition 1,  the minimum coding rate for a fixed $D$ corresponds to the rate-distortion function \cite[Sect.10]{m}. Thus, in the proof of achievability, we evaluate the coding rate and the distortion with the argument in rate-distortion theory. This view is also important to correctly understand the numerical results in Sect. \ref{sec:5}. \hspace{\fill}$\Box$
\end{rem}

\begin{dfn}
The closure of the set of achievable tuples $(R, D, E)$ is referred to as the {\bf achievable region} and is denoted by $\mathcal{C}(\mathcal{E})$.\hspace{\fill}$\Box$
\end{dfn}
The problems in the cases of $\mathcal{E} = \mathcal{K}$ and $\mathcal{E} = \mathcal{R}$ are firstly treated by Yamamoto \cite{a}. We show the results in Theorem \ref{thm:1}.
\begin{thm}[Yamamoto \cite{a}]

If $\mathcal{E} = \mathcal{K}$, the achievable region denoted as $\mathcal{C}_{Y}(\mathcal{K})$ is given by $\mathcal{S}_Y (\mathcal{K})$ such that
\begin{align}
\mathcal{S}_Y (\mathcal{K})=\{(R, D, E):~~
&R\ge I(X_{\mathcal{R}}, X_{\mathcal{H}};\hat{X}_{\mathcal{R}}),  \nonumber\\
&D\ge \mathbb{E}[d(X_{\mathcal{R}}, \hat{X}_{\mathcal{R}})], \nonumber\\
&0 \le E\le H(X_{\mathcal{H}}|\hat{X}_{\mathcal{R}})\nonumber\\
& \mathrm{for\ some}\ p(x_{\mathcal{R}}, x_{\mathcal{H}}, \hat{x}_{\mathcal{R}})\}.\nonumber
\end{align}
If $\mathcal{E} = \mathcal{R}$, the achievable region denoted as $\mathcal{C}_{Y}(\mathcal{R})$ is given by $\mathcal{S}_Y (\mathcal{R})$ such that
\begin{align}
\mathcal{S}_Y (\mathcal{R})=\{(R, D, E):~~
&R\ge I(X_{\mathcal{R}};\hat{X}_{\mathcal{R}}), \nonumber\\
&D\ge \mathbb{E}[d(X_{\mathcal{R}}, \hat{X}_{\mathcal{R}})], \nonumber\\
&0 \le E\le H(X_{\mathcal{H}}|\hat{X}_{\mathcal{R}})\nonumber\\
& \mathrm{for\ some}\ p(x_{\mathcal{R}}, x_{\mathcal{H}})p(\hat{x}_{\mathcal{R}}|x_{\mathcal{R}})\}.\nonumber\Box
\end{align}
\label{thm:1}
\end{thm}
\begin{rem}
Sankar et al. \cite{b} analyzed the coding systems in which the decoder can observe side information in addition to $X_{\mathcal{E}}^{n}$. The results in Theorem \ref{thm:1} can be viewed as a special case of the one in \cite{b} when the decoder cannot observe side information.\label{rem:1}\hspace{\fill}$\Box$
\end{rem}

\subsection{Main Theorem}

As we have mentioned before, in \cite{a}, \cite{b} the expressions of the achievable regions in the case of $\mathcal{E} = \mathcal{K}$ and $\mathcal{E} = \mathcal{R}$ are derived separately. However, if we express these regions in a unified way using $\mathcal{E}$ with $\mathcal{R} \subseteq \mathcal{E} \subseteq \mathcal{K}$, the expression may provide insights on how the region changes with $\mathcal{E}$. In this paper, we analyze the model with general $\mathcal{E} \supseteq \mathcal{R}$ and derive the unified expression of the achievable regions. We show the formula of the achievable region in Theorem \ref{thm:2}.
\begin{dfn}
For any $\mathcal{E}$ such that $\mathcal{R} \subseteq \mathcal{E} \subseteq \mathcal{K}$, $\mathcal{S}^{*}(\mathcal{E})$ is defined as
\begin{align}
\mathcal{S}^{*}(\mathcal{E})=\{(R, D, E):~~
&R\ge I(X_{\mathcal{E}};\hat{X}_{\mathcal{R}}), \nonumber\\
&D\ge \mathbb{E}[d(X_{\mathcal{R}}, \hat{X}_{\mathcal{R}})], \nonumber\\
&0 \le E\le H(X_{\mathcal{H}}|\hat{X}_{\mathcal{R}})\nonumber\\
& \mathrm{for\ some}\ p(x_{\mathcal{E}}, x_{\mathcal{E}^{\mathrm{c}}})p(\hat{x}_{\mathcal{R}}|x_{\mathcal{E}})\}. \label{eq:3.6}
\end{align}
\hspace{\fill}$\Box$\label{dfn:3}
\end{dfn}
\begin{thm}
For any $\mathcal{E}$ such that $\mathcal{R} \subseteq \mathcal{E} \subseteq \mathcal{K}$, the achievable region of the coding system is given by
\begin{align}
\mathcal{C}(\mathcal{E}) = \mathcal{S}^{*}(\mathcal{E}).\label{eq:3.7}
\end{align}
\hspace{\fill}$\Box$
\label{thm:2}
\end{thm}
\begin{rem}
The expression of the region in Theorem \ref{thm:2} unitedly represents various expressions of the regions corresponding to the set $\mathcal{E}$. For example, if $\mathcal{E} = \mathcal{R} \times \mathcal{H}$ or $\mathcal{E} = \mathcal{R}$, then \eqref{eq:3.7} reduces to two expressions of the regions $\mathcal{C}_{Y}^{1}$ and $\mathcal{C}_{Y}^{2}$ in Theorem \ref{thm:1} established in \cite{a}. \hspace{\fill}$\Box$ \label{rem:2}
\end{rem}

If the expression $\mathcal{S}^{*}(\mathcal{E})$ of the achievable region given in Theorem \ref{thm:2} is a convex set, we can calculate the boundary points of the achievable region effectively. Therefore, we briefly mention the convexity of the expression of the region. Indeed, we can prove that for any tuples $(R, D, E) \in \mathcal{S}^{*}(\mathcal{E})$, $(R', D', E') \in \mathcal{S}^{*}(\mathcal{E})$, and $\lambda \in (0, 1)$, there exists a conditional distribution $p_{\lambda}(\hat{x}_{\mathcal{R}} | x_{\mathcal{E}})$ such that
\begin{align}
\lambda R+(1-\lambda )R'&\ge I_{p_{\lambda}}(X_{\mathcal{E}};\hat{X}_{\mathcal{R}}), \nonumber\\
\lambda D+(1-\lambda )D'&\ge \mathbb{E}_{p_{\lambda}}[d(X_{\mathcal{R}}, \hat{X}_{\mathcal{R}})], \nonumber\\
\lambda E+(1-\lambda )E'&\le H_{p_{\lambda}}(X_{\mathcal{H}}|\hat{X}_{\mathcal{R}}).\nonumber
\end{align}
That is, $\mathcal{S}^{*}(\mathcal{E})$ is a convex set.

\section{Proof of Main Theorem (Theorem \ref{thm:2})}

\subsection{Proof Preliminaries}

For preliminaries, we define strongly typical sequences that are necessary for the proof and show some properties.
\begin{dfn}{\it \cite[Definition 2.1]{e}}

The type of a sequence $x^{n} \in \mathcal{X}^{n}$ of length $n$ is the distribution $P_{x^{n}}$ on $\mathcal{X}$ defined by
\begin{align}
P_{x^{n}}(a)\coloneqq\frac{1}{n}N(a|x^{n}), \label{eq:4.1}
\end{align}
where $N(a | x^{n})$ represents the number of the occurrences of symbol $a \in \mathcal{X}$ in $x^{n}$. Likewise, the joint type of $x^{n} \in \mathcal{X}^{n}$ and $y^{n} \in \mathcal{Y}^{n}$ is the distribution $P_{x^{n}y^{n}}$ on $\mathcal{X} \times \mathcal{Y}$ defined by
\begin{align}
P_{x^{n}y^{n}} \coloneqq \frac{1}{n}N(a, b | x^{n}, y^{n}),
\end{align}
where $N(a, b | x^{n}, y^{n})$ represents the number of the occurrences of $(a, b) \in \mathcal{X} \times \mathcal{Y}$ in the pair of sequences $(x^{n}, y^{n})$. 

\hspace{\fill}$\Box$\label{dfn:4}
\end{dfn}
\begin{dfn}{\it(Conditional Type) \cite[Definition 2.2]{e}}

We define the conditional type of $y^{n}$ given $x^{n}$ as a stochastic matrix $V\!:~\mathcal{X} \rightarrow \mathcal{Y}$ satisfying
\begin{align}
N(a, b|x^{n}, y^{n})=N(a|x^{n})V(b|a).\label{eq:4.2}
\end{align}
In particular, the conditional type of $y^{n}$ given $x^{n}$ is uniquely determined and given by
\begin{align}
V(b|a)=\frac{N(a, b|x^{n}, y^{n})}{N(a|x^{n})}\label{eq:4.3}
\end{align}
if $N(a | x^{n}) > 0$ for any $a \in \mathcal{X}$.\hspace{\fill}$\Box$\label{dfn:5}
\end{dfn}
\begin{dfn}{\it (Strongly Typical Sequences) \cite[Definition 1.2.8]{f}}

For any distribution $P$ on $\mathcal{X}$, a sequence $x^{n} \in \mathcal{X}^{n}$ is said to be $P$-typical with constant $\delta > 0$ if 
\begin{align}
\left| \frac{1}{n}N(a | x^{n}) - P(a) \right| \le \delta~~~~{\rm for~every~} a \in \mathcal{X} \label{eq:4.4}
\end{align}
and, in addition, no $a \in \mathcal{X}$ with $P(a) = 0$ occurs in $x^{n}$.
The set of such sequences is denoted by $T_{\delta}^{n}(P)$. If $X$ is a random variable with values in $\mathcal{X}$, we also refer to $P$-typical sequences as $X$-typical sequences and write $T_{\delta}^{n}(X)$.\hspace{\fill}$\Box$\label{dfn:6}
\end{dfn}
\begin{dfn}{\it (Conditional Strongly Typical Sequences) \cite[Definition 1.2.9]{f}}

For a stochastic matrix $W\!\!:~\mathcal{X}\rightarrow\mathcal{Y}$, a sequence $y^{n} \in \mathcal{Y}^{n}$ is said to be $W$-typical given $x^{n} \in \mathcal{X}^{n}$ with constant $\delta > 0$ if
\begin{align}
\left| \frac{1}{n}N(a, b | x^{n}, y^{n}) - \frac{1}{n}N(a | x^{n})W(b | a) \right| \le \delta&\label{eq:4.5}\\
{\rm for~every~}a \in \mathcal{X}, b \in \mathcal{Y},&\nonumber
\end{align}
and, in addition, $N(a, b | x^{n}, y^{n}) = 0$ whenever $W(b | a) = 0$. The set of such sequences $y^{n}$ is denoted by $T_{\delta}^{n}(W | x^{n})$. Further, if $X$ and $Y$ are random variables with values in $\mathcal{X}$ and $\mathcal{Y}$ respectively and $P_{Y|X} = W$, then they are also said to be $Y | X$-typical and written as $T_{\delta}^{n}(Y | X | x^{n})$.\hspace{\fill}$\Box$\label{dfn:7}
\end{dfn}
Hereafter, the set of conditional strongly typical sequences $T_{\delta}^{n}(Y | X | x^{n})$ is abbreviated as $T_{\delta}^{n}(Y | x^{n})$. 

We state some lemmas that are used in this proof. 
\begin{lem}{\it \cite[Lemma 1.2.13]{f}}\label{lem:6}

If $\delta_{n} \rightarrow 0$ and $\delta' \rightarrow 0$ as $n \rightarrow 0$, there exists a sequence $\epsilon_{n} = \epsilon_{n}(|\mathcal{X}, \mathcal{Y}|, \delta_{n}, \delta_{n}') \rightarrow 0$ $(n \rightarrow \infty)$ such that for every distribution $P$ on $\mathcal{X}$ and stochastic matrix $W\!\!:~\mathcal{X} \rightarrow \mathcal{Y}$
\begin{align}
\left|\frac{1}{n}\log{|T_{\delta_{n}}^{n}(P){|}-H(P)}\right|&\le\epsilon_{n},\\
\left|\frac{1}{n}\log{|T_{\delta'_{n}}^{n}(W|x^{n}){|}-H(W|P)}\right|&\le\epsilon_{n}.
\end{align}
\hspace{\fill}$\Box$\label{thm:3}
\end{lem}
\begin{lem}{\it \cite[Lemma 1.2.7]{f}}

Let the variational distance between two distributions $P$ and $Q$ on $\mathcal{X}$ be defined as
\begin{align}
d_{{\rm v}}(P, Q) \coloneqq \sum_{x \in \mathcal{X}} |P(x) - Q(x)|.
\end{align}
If $d_{{\rm v}}(P, Q) < \frac{1}{2}$, then
\begin{align}
|H(P) - H(Q)| \le -d_{{\rm v}}(P, Q) \cdot \log{\frac{d_{{\rm v}}(P, Q)}{|\mathcal{X}|}}.
\end{align}
\hspace{\fill}$\Box$\label{lem:1}
\end{lem}
\begin{lem}{\it \cite[Lemma 1.2.10]{f}}

If $x^{n} \in T_{\delta}^{n}(X)$ and $y^{n} \in T_{\delta'}^{n}(Y | x^{n})$, then $(x^{n}, y^{n}) \in T_{\delta + \delta'}^{n}(X, Y)$ and, consequently, $y^{n} \in T_{\delta''}(Y)$ for $\delta'' \coloneqq (\delta + \delta')\cdot|\mathcal{X}|$.\hspace{\fill}$\Box$\label{lem:2}
\end{lem}
\begin{lem}

If $(x^{n}, y^{n}) \in T_{\delta}^{n}(X, Y)$, then $x^{n} \in T_{\delta_{1}}^{n}(X)$ and, consequently, $y^{n} \in T_{\delta_{2}}^{n}(Y | x^{n})$ for $\delta_{1} \coloneqq |\mathcal{Y}| \cdot \delta$ and $\delta_{2} \coloneqq (|\mathcal{Y}| + 1) \cdot \delta$.

\hspace{\fill}$\Box$\label{lem:3}
\end{lem}
\begin{lem}

If $y^{n} \in T_{\delta}^{n}(Y)$ and $(x^{n}, y^{n}) \notin T_{2\delta}^{n}(X, Y)$, then $x^{n} \notin T_{\delta}^{n}(X | y^{n})$.\hspace{\fill}$\Box$\label{lem:4}
\end{lem}
\begin{lem}
\rm{{\cite[Lemma 1.2.12 and Remark]{f}}}

For arbitrarily fixed $\delta > 0$ and every distribution $P$ on $\mathcal{X}$ and stochastic matrix $W\!\!:~\mathcal{X} \rightarrow \mathcal{Y}$
\begin{align}
\Pr\{X^{n} \in T_{\delta}^{n}(P)\}&\ge 1-2|\mathcal{X}|\mathrm{e}^{-2\delta^{2}n},\\
\Pr\{Y^{n} \in T_{\delta}^{n}(W|x^{n}) | X^{n} = x^{n}\} &\ge 1-2|\mathcal{X}| \cdot |\mathcal{Y}|\mathrm{e}^{-2\delta^{2}n}\nonumber\\
&\hspace{1.1mm}\mathrm{for~every}~x^{n} \in \mathcal{X}^{n}. 
\end{align}
\hspace{\fill}$\Box$\label{lem:5}
\end{lem}

\subsection{Proof of Converse Part}

In this part, we shall prove $\mathcal{C}(\mathcal{E})\subseteq\mathcal{S}^{*}(\mathcal{E})$.

Let a tuple $(R, D, E)\in\mathcal{C}(\mathcal{E})$ be arbitrarily fixed. Then, there exists an $(n, 2^{n(R+\epsilon)}, D+\epsilon, E-\epsilon )$ code that satisfies (\ref{eq:3.1})--(\ref{eq:3.3}). Let $Q$ be a uniform random variable over $\{1, 2, \ldots, n\}$ and let $p_{i}(x_{\mathcal{E}, i}, x_{\mathcal{E}^{\mathrm{c}}, i}, \hat{x}_{\mathcal{R}, i})$ be the conditional distribution given $Q = i$. Evaluating the inequalities for $R$, we obtain
\begin{align}
R+\epsilon&\stackrel{\rm{(a)}}{\ge} \frac{1}{n}\log{M_{n}}\nonumber\\
&\stackrel{\rm{(b)}}{\ge} \frac{1}{n}H(J_{n})\nonumber\\
&\ge \frac{1}{n}I(J_{n};X_{\mathcal{E}}^{n})\nonumber\\
&\stackrel{\rm{(c)}}{=} \frac{1}{n}\{H(X_{\mathcal{E}}^{n}) - H(X_{\mathcal{E}}^{n}|J_{n}, \hat{X}_{\mathcal{R}}^{n})\}\nonumber\\
&\stackrel{\rm{(d)}}{=} \frac{1}{n}\sum_{i=1}^{n}H(X_{\mathcal{E}, i})-\frac{1}{n}\sum_{i=1}^{n}H(X_{\mathcal{E}, i}|X_{\mathcal{E}}^{i-1}, J_{n}, \hat{X}_{\mathcal{R}}^{n})\nonumber\\
&\stackrel{\rm{(e)}}{\ge}\frac{1}{n}\sum_{i=1}^{n}H(X_{\mathcal{E}, i})-\frac{1}{n}\sum_{i=1}^{n}H(X_{\mathcal{E}, i}|\hat{X}_{\mathcal{R}, i})\nonumber\\
&\stackrel{\rm{(f)}}{=} \sum_{i=1}^{n}\Pr\{Q=i\}H(X_{\mathcal{E}, i}|Q=i)\nonumber\\
&{\hspace{4.5mm}-}\sum_{i=1}^{n}\Pr\{Q=i\}H(X_{\mathcal{E}, i}|\hat{X}_{\mathcal{R}, i},Q=i)\nonumber\\
&= H(X_{\mathcal{E}, Q}|Q)-H(X_{\mathcal{E}, Q}|\hat{X}_{\mathcal{R}, Q}, Q)\nonumber\\
&\stackrel{\rm{(g)}}{=} H(X_{\mathcal{E}})-H(X_{\mathcal{E}, Q}|\hat{X}_{\mathcal{R}, Q}, Q)\nonumber\\
&\stackrel{\rm{(h)}}{\ge} H(X_{\mathcal{E}})-H(X_{\mathcal{E}}|\hat{X}_{\mathcal{R}})\nonumber\\
&=I(X_{\mathcal{E}}; \hat{X}_{\mathcal{R}}), \label{eq:4.7}
\end{align}
where
\begin{itemize}
\item[(a)] follows from \eqref{eq:3.1},
\item[(b)] follows because $H(J_{n}) \le \log{|J_{n}|} = \log{M_{n}}$,
\item[(c)] is due to the fact that $\hat{X}_{\mathcal{R}}^{n} = g(J_{n})$,
\item[(d)] follows because each $X_{\mathcal{K}, i}$ is independent and $\hat{X}_{\mathcal{R}}^{n}$ is a function of $J_{n}$,
\item[(e)] follows because conditioning reduces entropy,
\item[(f)] is due to the definition of $Q$,
\item[(g)] follows because $X_{\mathcal{E}} \perp Q$, 
\item[(h)] follows because conditioning reduces entropy, where  $(X_{\mathcal{E}}, \hat{X}_{\mathcal{R}}) \sim \sum_{i=1}^{n}\Pr\{Q=i\}p_{i}(x_{\mathcal{E}, i}, \hat{x}_{\mathcal{R}, i}) = p(x_{\mathcal{E}}, \hat{x}_{\mathcal{R}})$.
\end{itemize}

Similarly, evaluating $D$ and $E$, respectively, we obtain
\begin{align}
D+\epsilon
&\stackrel{\rm{(i)}}{\ge} \mathbb{E}\left[\frac{1}{n}\sum_{i=1}^{n}d(X_{\mathcal{R}, i}, \hat{X}_{\mathcal{R}, i})\right]\nonumber\\
&= \frac{1}{n}\sum_{i=1}^{n}\mathbb{E}[d(X_{\mathcal{R}, i}, \hat{X}_{\mathcal{R}, i})]\nonumber\\
&\stackrel{\rm{(j)}}{=} \mathbb{E}_{Q}[\mathbb{E}[d(X_{\mathcal{R}, i}, \hat{X}_{\mathcal{R}, i})|Q]]\nonumber\\
&\stackrel{\rm{(k)}}{=}\mathbb{E}[d(X_{\mathcal{R}}, \hat{X}_{\mathcal{R}})],\label{eq:4.8}\\
E-\epsilon
&\stackrel{\rm{(l)}}{\le}\frac{1}{n}H(X_{\mathcal{H}}^{n}|J_{n})\nonumber\\
&= \frac{1}{n}\sum_{i=1}^{n}H(X_{\mathcal{H}, i}|X_{\mathcal{H}}^{i-1}, J_{n})\nonumber\\
&\stackrel{\rm{(m)}}{=} \frac{1}{n}\sum_{i=1}^{n}H(X_{\mathcal{H}, i}|X_{\mathcal{H}}^{i-1}, J_{n}, \hat{X}_{\mathcal{R}, i})\nonumber\\
&\stackrel{\rm{(n)}}{\le}\frac{1}{n}\sum_{i=1}^{n}H(X_{\mathcal{H}, i}|\hat{X}_{\mathcal{R}, i})\nonumber\\
&\stackrel{\rm{(o)}}{=}\sum_{i=1}^{n}\Pr\{Q=i\}H(X_{\mathcal{H}, i}|\hat{X}_{\mathcal{R}, i}, {Q = i})\nonumber\\
&=H(X_{\mathcal{H}, Q}|\hat{X}_{\mathcal{R}, Q}, Q)\nonumber\\
&\stackrel{\rm{(p)}}{\le} H(X_{\mathcal{H}}|\hat{X}_{\mathcal{R}}), \label{eq:4.9}
\end{align}
where
\begin{itemize}
\item[(i)] is due to \eqref{eq:3.2},
\item[(j)] is derived from the definition of $Q$,
\item[(k)] follows because $(X_{\mathcal{R}}, \hat{X}_{\mathcal{R}})\sim\sum_{i=1}^{n}\Pr\{Q=i\}p_{i}(x_{\mathcal{R}, i}, \hat{x}_{\mathcal{R}, i})=p(x_{\mathcal{R}}, \hat{x}_{\mathcal{R}})$,
\item[(l)] is due to \eqref{eq:3.3},
\item[(m)] follows because $\hat{X}_{\mathcal{R}}^{n} = g(J_{n})$,
\item[(n)] follows from the fact that conditioning reduces entropy,
\item[(o)] is derived from the definition of $Q$,
\item[(p)] follows because conditioning reduces entropy, where $(X_{\mathcal{H}}, \hat{X}_{\mathcal{R}}) \sim \sum_{i=1}^{n}\Pr\{Q=i\}p_{i}(x_{\mathcal{H}, i}, \hat{x}_{\mathcal{R}, i})=p(x_{\mathcal{H}}, \hat{x}_{\mathcal{R}})$.
\end{itemize}
It is readily shown that the Markov chain $X_{\mathcal{E}^{\mathrm{c}}}$--$X_{\mathcal{E}}$--$\hat{X}_{\mathcal{R}}$ holds (Appendix \ref{apt:A}). We complete the proof of the converse part.\hspace{\fill}$\Box$

\subsection{Proof of Direct Part (Achievability)}

In this part, we provide a sketch of the proof of $\mathcal{S}^{*}(\mathcal{E})\subseteq\mathcal{C}(\mathcal{E})$.

Under an arbitrarily fixed distribution $p(x_{\mathcal{E}},x_{\mathcal{E}^{\mathrm{c}}})p(\hat{x}_{\mathcal{R}}|x_{\mathcal{E}})$, any tuple $(R, D, E)\in\mathcal{S}^{*}(\mathcal{E})$ is chosen such that
\begin{align}
R&>I(X_{\mathcal{E}};\hat{X}_{\mathcal{R}}), \label{eq:4.10}\\
D&>\mathbb{E}[d(X_{\mathcal{R}}, \hat{X}_{\mathcal{R}})], \label{eq:4.11}\\
E&<H(X_{\mathcal{H}}|\hat{X}_{\mathcal{R}}). \label{eq:4.12}
\end{align}
From $\eqref{eq:4.11}$ and $\eqref{eq:4.12}$, we can choose a sufficiently small $\epsilon > 0$ such that
\begin{align}
D&>\mathbb{E}[d(X_{\mathcal{R}}, \hat{X}_{\mathcal{R}})]+\epsilon, \label{eq:4.13}\\
E&<H(X_{\mathcal{H}}|\hat{X}_{\mathcal{R}})-\epsilon. \label{eq:4.14}
\end{align}
In addition, with this $\epsilon$, some constant $0 < \tau < \frac{1}{2}$ is fixed such that
\begin{align}
&\tau (\log{|\mathcal{X}_{\mathcal{H}}|}+5) + 4\tau\log\frac{|\mathcal{X}_{\mathcal{H}}| \cdot 2^{R}}{2\tau} < \epsilon. \label{eq:4.15}
\end{align}
Also, we can choose positive numbers $\delta(\coloneqq\delta(n))$ such that
\begin{align}
&(\delta(n) + \delta_{1}(n))|\mathcal{X}_{\mathcal{R}}| {\cdot} |\hat{\mathcal{X}}_{\mathcal{R}}| D_{\max}+\tau<\epsilon, \label{eq:4.16}\\
&2\delta^{2}(n) \le R - I(X_{\mathcal{E}} ; \hat{X}_{\mathcal{R}}) - \frac{1}{n} - \tau, \label{eq:4.17}\\
&\delta (n)\rightarrow 0, \label{eq:4.18}\\
&\sqrt{n}\cdot\delta (n)\rightarrow\infty \label{eq:4.19}
\end{align}
as $n \rightarrow \infty$, where $\delta_{1} \coloneqq (|\mathcal{X}_{\mathcal{E}}| - |\mathcal{X}_{\mathcal{R}}|) \cdot \delta$ and $D_{\max}\coloneqq \underset{a\in\mathcal{X}_{\mathcal{R}}, b\in\hat{\mathcal{X}}_{\mathcal{R}}}{\max}d(a, b)$. Let $\delta(n) = \frac{c}{\sqrt{n}}\log{n}$ where $c$ is a constant, and obviously \eqref{eq:4.18} and \eqref{eq:4.19} are satisfied.

{\bf Generation of codebook}:\quad Randomly generate $\hat{x}_{\mathcal{R}}^{n}(j)$ from the strongly typical sequences $T_{\delta}^{n}(\hat{X}_{\mathcal{R}})$ for $j=1, {2}, \ldots, M_{n} \coloneqq 2^{nR}$. Reveal the codebook $\mathcal{C}=\{\hat{x}_{\mathcal{R}}^{n}(1), \ldots, \hat{x}_{\mathcal{R}}^{n}(M_{n})\}$ to the encoder and decoder.

{\bf Encoding}:\quad If a sequence $x_{\mathcal{E}}^{n} \in \mathcal{X}_{\mathcal{E}}^{n}$ satisfies $x_{\mathcal{K}}^{n} = (x_{\mathcal{E}}^{n}, x_{\mathcal{E}^{\mathrm{c}}}^{n})$ with some $x_{\mathcal{E}^{\mathrm{c}}}^{n} \in \mathcal{X}_{\mathcal{E}^{\mathrm{c}}}^{n}$, we write $x_{\mathcal{E}}^{n}\prec x_{\mathcal{K}}^{n}$. Given $x_{\mathcal{K}}^{n}$, the encoder finds $j$ such that $x_{\mathcal{E}}^{n} \in T_{\delta}^{n}(X_{\mathcal{E}} | \hat{x}_{\mathcal{R}}(j))$ and sets $f_{n}(x_{\mathcal{E}}^{n}) = j$ where $T_{\delta}^{n}(X_{\mathcal{E}} | \hat{x}_{\mathcal{R}}(j))$ is the conditional strongly typical sequences. If there exist multiple such $j$, $f_{n}(x_{\mathcal{E}}^{n})$ is set as the minimum one.  If there are no such $j$, then $f_{n}(x_{\mathcal{E}}^{n}) = M_{n}$.

{\bf Decoding}:\quad When $j$ is observed, the decoder sets the reproduced sequence as $\hat{X}_{\mathcal{R}}^{n} = \hat{x}_{\mathcal{R}}^{n}(j)$.

{\bf Evaluation}:\quad We define $\mathcal{A}(j)$, $\mathcal{B}(j)$, and $\tilde{\mathcal{A}}(j)$ as
\begin{align}
\mathcal{A}(j)&\coloneqq\{x_{\mathcal{E}}^{n}\!:~f_{n}(x_{\mathcal{E}}^{n})=j\}, \label{eq:4.20}\\
\mathcal{B}(j)&\coloneqq\{x_{\mathcal{K}}^{n}\!:~x_{\mathcal{E}}^{n}\prec x_{\mathcal{K}}^{n}, f_{n}(x_{\mathcal{E}}^{n})=j\}, \label{eq:4.21}\\
\tilde{\mathcal{A}}(j)&\coloneqq
\begin{cases}
\{x_{\mathcal{K}}^{n}\!:~x_{\mathcal{E}}^{n}\prec x_{\mathcal{K}}^{n}, f_{n}(x_{\mathcal{E}}^{n})=j, x_{\mathcal{K}}^{n}\in T_{2\delta}^{n}(X_{\mathcal{K}} | \hat{x}_{\mathcal{R}}^{n}(j))\}\\
\hspace{4.2cm}(j = 1, {2}, \ldots, M_{n}-1)\\
\{x_{\mathcal{K}}^{n}\!:~x_{\mathcal{K}}^{n} \in \mathcal{X}_{\mathcal{K}}^{n} \setminus \bigcup_{j = 1}^{M_{n} - 1} \mathcal{\tilde{A}}(j)\}~~~~(j = M_{n})
\end{cases} \label{eq:4.22}
\end{align}
It is easily verified that $\mathcal{A}(j)$ for $j = 1, 2, \ldots, M_{n}$ (also, $\mathcal{B}(j)$ and $\tilde{\mathcal{A}}(j)$) are disjoint. From the definitions of $J_{n}$, $\mathcal{A}(j)$, and $\mathcal{B}(j)$,
\begin{align}
\Pr\{J_{n}=j\}=\Pr\{X_{\mathcal{E}}^{n}\in\mathcal{A}(j)\}=\Pr\{X_{\mathcal{K}}^{n}\in\mathcal{B}(j)\}&\label{eq:56}\\
\mathrm{for~} j=1, 2, \ldots, M_{n}.&\nonumber
\end{align}
For sufficiently large $n$, we can prove (Appendix \ref{apx:B})
\begin{align}
|\Pr\{X_{\mathcal{K}}^{n} \in \mathcal{B}(j)\} - \Pr\{X_{\mathcal{K}}^{n} \in \tilde{\mathcal{A}}(j)\}| \le 2|\mathcal{X}_{\mathcal{K}}|\cdot|\hat{\mathcal{X}}_{\mathcal{R}}|\mathrm{e}^{-2\delta^{2}n}&\label{eq:3.50}\\
\mathrm{for~}j = 1, 2, \ldots, M_{n} -1.& \nonumber
\end{align}

For sufficiently large $n$, we can show that there exists a code $(f_{n}, g_{n})$ such that (Appendix \ref{apx:C})
\begin{align}
&r_{n}\le R, \label{eq:4.25}\\
&u_{n}\le\mathbb{E}[d(X_{\mathcal{R}}, \hat{X}_{\mathcal{R}})]+(\delta + \delta_{1}) |\mathcal{X}_{\mathcal{R}}| \cdot |\hat{\mathcal{X}}_{\mathcal{R}}|D_{\max}+\tau, \label{eq:4.26}\\
&\Pr\left\{X_{\mathcal{E}}^{n} \notin \bigcup_{j=1}^{M_{n}-1}\mathcal{A}(j)\right\} \le (2|\mathcal{X}_{\mathcal{E}}| + 1)\mathrm{e}^{-2\delta^{2}n}, \label{eq:4.27}\\
&\Pr\left\{X_{\mathcal{K}}^{n} \notin \bigcup_{j=1}^{M_{n}-1}\tilde{\mathcal{A}}(j)\right\}\le\tau, \label{eq:4.28}\\
&|\tilde{\mathcal{A}}(j)|\ge 2^{n\{H(X_{\mathcal{K}}|\hat{X}_{\mathcal{R}})-\tau\}}. \label{eq:4.29}
\end{align}
For this code $(f_{n}, g_{n})$, we evaluate the privacy as
\begin{align}
e_{n}&\coloneqq\frac{1}{n}H(X_{\mathcal{H}}^{n}|J_{n})\nonumber\\
&= \frac{1}{n}\sum_{j=1}^{M_{n}}H(X_{\mathcal{H}}^{n}|X_{\mathcal{K}}^{n} \in \mathcal{B}(j))\Pr\{X_{\mathcal{K}}^{n} \in \mathcal{B}(j)\}\nonumber\\
&\stackrel{\rm{(a)}}{>}\frac{1}{n}\sum_{j=1}^{M_{n}}H(X_{\mathcal{H}}^{n}|X_{\mathcal{K}}^{n}\in \tilde{\mathcal{A}}(j))\Pr\{X_{\mathcal{K}}^{n}\in\tilde{\mathcal{A}}(j)\}\nonumber\\
&\hspace{4.5mm} - 4\tau\log\frac{|\mathcal{X}_{\mathcal{H}}| \cdot 2^{R}}{2\tau} \label{eq:3.62}\\
&\stackrel{\rm{(b)}}{>}\frac{1}{n}\sum_{j=1}^{M_{n}-1}H(X_{\mathcal{H}}^{n}|X_{\mathcal{K}}^{n}\in \tilde{\mathcal{A}}(j))\Pr\{X_{\mathcal{K}}^{n}\in\tilde{\mathcal{A}}(j)\} \nonumber\\
&\hspace{4.5mm}- 4\tau\log\frac{|\mathcal{X}_{\mathcal{H}}| \cdot 2^{R}}{2\tau} \nonumber\\
&=\frac{1}{n}\sum_{j=1}^{M_{n}-1}\bigg{[}-\sum_{\substack{x_{\mathcal{H}}^{n}}}\Pr\{X_{\mathcal{H}}^{n}=x_{\mathcal{H}}^{n}|X_{\mathcal{K}}^{n}\in\tilde{\mathcal{A}}(j)\}\cdot\nonumber\\
&\hspace{4.5mm}\log{\Pr\{X_{\mathcal{H}}^{n}=x_{\mathcal{H}}^{n}|X_{\mathcal{K}}^{n}\in\tilde{\mathcal{A}}(j)\}}\bigg{]}\cdot\nonumber\\
&\hspace{4.5mm}\Pr\{X_{\mathcal{K}}^{n}\in\tilde{\mathcal{A}}(j)\} - 4\tau\log\frac{|\mathcal{X}_{\mathcal{H}}| \cdot 2^{R}}{2\tau},\label{eq:54}
\end{align}
where
\begin{itemize}
\item[(a)] is due to the inequality proved in Appendix D,
\item[(b)] follows by removing the term for $j = M_{n}$.
\end{itemize}
Here, for any $x_{\mathcal{H}}^{n}$ satisfying $x_{\mathcal{K}}^{n} = (x_{\mathcal{R}}^{n}, x_{\mathcal{H}}^{n})\in\tilde{\mathcal{A}}(j)$ with some $x_{\mathcal{R}}^{n}$, we can show that
\begin{align}
&\Pr\{X_{\mathcal{H}}^{n}=x_{\mathcal{H}}^{n}|X_{\mathcal{K}}^{n}\in\tilde{\mathcal{A}}(j)\}\nonumber\\
&=\frac{\Pr\{X_{\mathcal{K}}^{n}\in\tilde{\mathcal{A}}(j)|X_{\mathcal{H}}^{n}=x_{\mathcal{H}}^{n}\}\Pr\{X_{\mathcal{H}}^{n}=x_{\mathcal{H}}^{n}\}}{\Pr\{X_{\mathcal{K}}^{n}\in\tilde{\mathcal{A}}(j)\}}\nonumber\\
&=\frac{\sum_{x_{\mathcal{R}}^{n}\!:~(x_{\mathcal{R}}^{n}, x_{\mathcal{H}}^{n})\in\tilde{\mathcal{A}}(j)}\Pr\{X_{\mathcal{R}}^{n}=x_{\mathcal{R}}^{n}, X_{\mathcal{H}}^{n}=x_{\mathcal{H}}^{n}|X_{\mathcal{H}}^{n}=x_{\mathcal{H}}^{n}\}}{\sum_{(\tilde{x}_{\mathcal{R}}^{n}, \tilde{x}_{\mathcal{H}}^{n})\in\tilde{\mathcal{A}}(j)}\Pr\{X_{\mathcal{R}}^{n}=\tilde{x}_{\mathcal{R}}^{n}, X_{\mathcal{H}}^{n}=\tilde{x}_{\mathcal{H}}^{n}\}}\cdot\nonumber\\
&\hspace{4.5mm}\Pr\{X_{\mathcal{H}}^{n}=x_{\mathcal{H}}^{n}\} \nonumber\\
&\stackrel{\rm{(c)}}{=}\frac{\sum_{x_{\mathcal{R}}^{n}\!:~(x_{\mathcal{R}}^{n}, x_{\mathcal{H}}^{n})\in\tilde{\mathcal{A}}(j)}\Pr\{X_{\mathcal{R}}^{n}=x_{\mathcal{R}}^{n}|X_{\mathcal{H}}^{n}=x_{\mathcal{H}}^{n}\}}{\sum_{(\tilde{x}_{\mathcal{R}}^{n}, \tilde{x}_{\mathcal{H}}^{n})\in\tilde{\mathcal{A}}(j)}\Pr\{X_{\mathcal{R}}^{n}=\tilde{x}_{\mathcal{R}}^{n}, X_{\mathcal{H}}^{n}=\tilde{x}_{\mathcal{H}}^{n}\}}\cdot\nonumber\\
&\hspace{4.5mm}\Pr\{X_{\mathcal{H}}^{n}=x_{\mathcal{H}}^{n}\}\nonumber\\
&\stackrel{\rm{(d)}}{\le}\frac{\sum_{x_{\mathcal{R}}^{n}\in T_{\delta_{3}}^{n}(X_{\mathcal{R}}|x_{\mathcal{H}}^{n}, \hat{x}_{\mathcal{R}}^{n}(j))}\Pr\{X_{\mathcal{R}}^{n}=x_{\mathcal{R}}^{n}|X_{\mathcal{H}}^{n}=x_{\mathcal{H}}^{n}\}}{\sum_{(\tilde{x}_{\mathcal{R}}^{n}, \tilde{x}_{\mathcal{H}}^{n})\in\tilde{\mathcal{A}}(j)}\Pr\{X_{\mathcal{R}}^{n}=\tilde{x}_{\mathcal{R}}^{n}, X_{\mathcal{H}}^{n}=\tilde{x}_{\mathcal{H}}^{n}\}}\cdot\nonumber\\
&\hspace{4.5mm}\Pr\{X_{\mathcal{H}}^{n}=x_{\mathcal{H}}^{n}\}\label{eq:3.68}\\
&\stackrel{\rm{(e)}}{\le}\frac{2^{n\{H(X_{\mathcal{R}}|X_{\mathcal{H}}, \hat{X}_{\mathcal{R}})+\tau\}}\cdot 2^{-n\{H(X_{\mathcal{R}}|X_{\mathcal{H}})-\tau\}}}{2^{n\{H(X_{\mathcal{K}}|\hat{X}_{\mathcal{R}})-\tau\}}\cdot2^{-n\{H(X_{\mathcal{K}})+\tau\}}}\cdot\nonumber\\
&\hspace{5.3mm}2^{-n\{H(X_{\mathcal{H}})-\tau\}}\nonumber\\
&=2^{-n\{H(X_{\mathcal{H}}|\hat{X}_{\mathcal{R}})-5\tau\}},\label{eq:59}
\end{align}
where
\begin{itemize}
\item[(c)] follows from the fact that 
\begin{align}
&\Pr\{X_{\mathcal{R}}^{n}=x_{\mathcal{R}}^{n}, X_{\mathcal{H}}^{n}=x_{\mathcal{H}}^{n}|X_{\mathcal{H}}^{n}=x_{\mathcal{H}}^{n}\}\nonumber\\
&=\Pr\{X_{\mathcal{R}}^{n}=x_{\mathcal{R}}^{n}|X_{\mathcal{H}}^{n}=x_{\mathcal{H}}^{n}\},\nonumber
\end{align}
\item[(d)] is due to the inequality proved in Appendix E,
\item[(e)] follows because the size of strongly typical sequences.
\end{itemize}
Therefore, from equations \eqref{eq:4.28}, \eqref{eq:54}, and \eqref{eq:59} we can obtain
\begin{align}
e_{n}
&> \frac{1}{n}\sum_{j=1}^{M_{n}-1}\bigg{[}n\sum_{\substack{x_{\mathcal{H}}^{n}}}\Pr\{X_{\mathcal{H}}^{n}=x_{\mathcal{H}}^{n}|X_{\mathcal{K}}^{n}\in\tilde{\mathcal{A}}(j)\}\cdot\nonumber\\
&\hspace{4.5mm}\{H(X_{\mathcal{H}}|\hat{X}_{\mathcal{R}})-5\tau\}\bigg{]}\cdot \Pr\{X_{\mathcal{K}}^{n}\in\tilde{\mathcal{A}}(j)\}\nonumber\\
&\hspace{4.5mm}- 4\tau\log\frac{|\mathcal{X}_{\mathcal{H}}| \cdot 2^{R}}{2\tau}\nonumber\\
&=\Pr\left\{X_{\mathcal{K}}^{n}\in\left(\bigcup_{j=1}^{M_{n}-1}\tilde{\mathcal{A}}(j)\right)\right\}\cdot\{H(X_{\mathcal{H}}|\hat{X}_{\mathcal{R}})-5\tau\}\nonumber\\
&\hspace{4.5mm} - 4\tau\log\frac{|\mathcal{X}_{\mathcal{H}}| \cdot 2^{R}}{2\tau}\nonumber\\
&\ge (1-\tau )\{H(X_{\mathcal{H}}|\hat{X}_{\mathcal{R}})-5\tau\} - 4\tau\log\frac{|\mathcal{X}_{\mathcal{H}}| \cdot 2^{R}}{2\tau}.\label{eq:4.32}
\end{align}
~~Since constants $\epsilon$, $\tau$, and $\delta$ are fixed to satisfy \eqref{eq:4.13}--\eqref{eq:4.16}, from \eqref{eq:4.25}, \eqref{eq:4.26}, and \eqref{eq:4.32}, we obtain
\begin{align}
r_{n}&\le R, \label{eq:4.33}\\
u_{n}&\le\mathbb{E}[d(X_{\mathcal{R}},\hat{X}_{\mathcal{R}})]+\epsilon<D, \label{eq4.34}\\
e_{n}&> H(X_{\mathcal{H}}|\hat{X}_{\mathcal{R}})-\epsilon>E. \label{eq:4.35}
\end{align}
Therefore, for the fixed distribution $p(x_{\mathcal{E}}, x_{\mathcal{E}^{\mathrm{c}}})p(\hat{x}_{\mathcal{R}}|x_{\mathcal{E}})$ any tuple
\begin{align}
(R, D, E)\in\{(R, D, E):~~
R&>I(X_{\mathcal{E}};\hat{X}_{\mathcal{R}}), \nonumber\\
D&>\mathbb{E}[d(X_{\mathcal{R}}, \hat{X}_{\mathcal{R}})], \nonumber\\
E&<H(X_{\mathcal{H}}|\hat{X}_{\mathcal{R}})\}\eqqcolon\mathcal{S}^{*}_{p}(\mathcal{E})\label{eq:4.36}
\end{align}
is achievable. Consequently, $\mathcal{S}^{*}_{p}(\mathcal{E})\subseteq\mathcal{C}(\mathcal{E})$. Taking the closure for the left-hand side, we obtain $Cl(\mathcal{S}^{*}_{p}(\mathcal{E}))\subseteq\mathcal{C}(\mathcal{E})$ because $\mathcal{C}(\mathcal{E})$ is a closed set. We conclude that $\mathcal{S}^{*}(\mathcal{E})=\bigcup_{p}Cl(\mathcal{S}^{*}_{p}(\mathcal{E}))\subseteq\mathcal{C}(\mathcal{E})$ because the distribution $p=p(x_{\mathcal{E}}, x_{\mathcal{E}^{\mathrm{c}}})p(\hat{x}_{\mathcal{R}}|x_{\mathcal{E}})$ is fixed arbitrarily. We complete the proof of the direct part.\hspace{\fill}$\Box$

\newpage
\section{Numerical Calculation}\label{sec:5}

In general, it is difficult to compute the achievable region $\mathcal{C}(\mathcal{E})$ in Theorem 2. Nevertheless, to get some insight, let us consider the three tractable but essential cases. In these calculations, the number of public attributes is one $(|\mathcal{R}|=1)$ and the number of private attributes is two $(|\mathcal{H}|=2)$. We assume that each of attributes is binary.  The measure of privacy given by \eqref{eq:2.12} can also be used to measure ``the amount of privacy leakage" by $l_{n} = H(X_{\mathcal{H}}) - e_{n}$. By using $l_{n}$ instead of $e_{n}$, the privacy condition \eqref{eq:3.3}, is replaced by $l_{n} \le L + \epsilon$, and the achievable region denoted by $\mathcal{C}_{RDL}$ is given by replacing $0 \le E \le H(X_{\mathcal{H}} | \hat{X}_{\mathcal{R}})$ with $L \ge I(X_{\mathcal{H}} ; \hat{X}_{\mathcal{R}})$ in (\ref{eq:3.6}). Here, note again that the coding rate $R$ acts like the rate-distortion function in rate-distortion theory ([12, Sect.10]).  For fixed $D$ and $L$, a smaller coding rate is better.

\begin{figure}[htbp]
    \centering
    \includegraphics[keepaspectratio, scale=0.2]{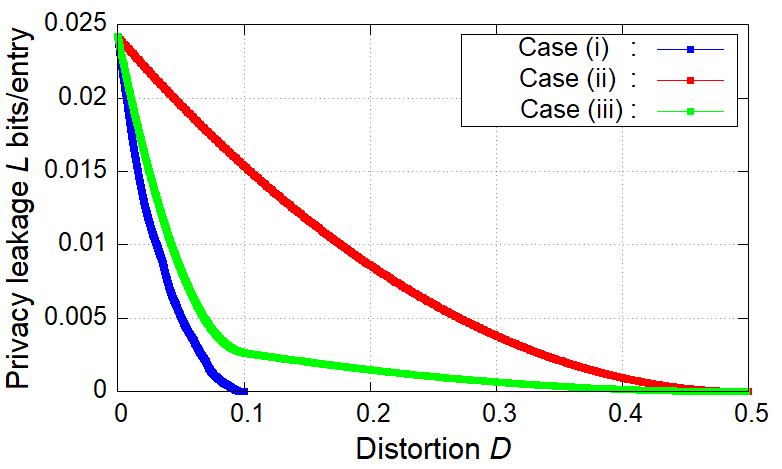}
    \vspace{-0.3cm}
    \caption{Utility-Privacy Trade-off Region in Cases (i), (ii), and (iii).}
    \label{fig:2}
    \centering
    \vspace{0.2cm}
    \includegraphics[keepaspectratio, scale=0.2]{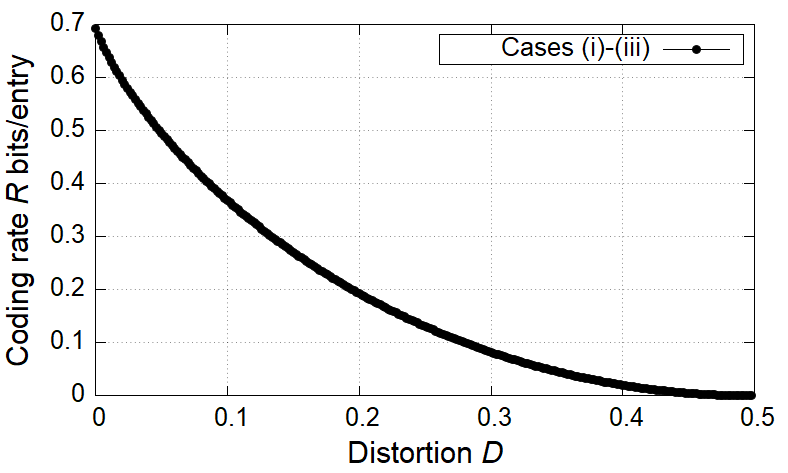}
    \vspace{-0.3cm}
    \caption{Utility-Coding Rate Trade-off Region in Cases (i), (ii), and (iii). The curves coincide in all cases.}
    \label{fig:3}
\end{figure}

\begin{table}[htbp]
\vspace{-0.2cm}
    \begin{center}
      \caption{Minimum $L$ and its corresponding $R$ \\for $D = 0.0500$}
      \label{tbl:1}
      \vspace{-0.3cm}
      \scalebox{0.92}{
\begin{tabular}{|c|c|c|} \hline
    Cases & Leakage $L$ & Coding rate $R$\\ \hline \hline
    \multicolumn{1}{|l|}{case (ii)} & 0.019512 & 0.494629 \\
    \multicolumn{1}{|l|}{case (iii)} & 0.008298 & 0.527700 \\
    \multicolumn{1}{|l|}{case (i)} & 0.005107 & 0.539478 \\ \hline
\end{tabular}}
    \end{center}
    \vspace{-0.1cm}
    \begin{center}
      \caption{Minimum $L$ and its corresponding $R$ \\for $D = 0.100$}
      \label{tbl:2}
      \vspace{-0.3cm}
      \scalebox{0.92}{
\begin{tabular}{|c|c|c|} \hline
    Cases & Leakage $L$ & Coding rate $R$\\ \hline \hline
    \multicolumn{1}{|l|}{case (ii)} & 0.015378 & 0.368062 \\
    \multicolumn{1}{|l|}{case (iii)} & 0.002656 & 0.418826 \\
    \multicolumn{1}{|l|}{case (i)} & 0.000000 & 0.429490 \\ \hline
\end{tabular}}
    \end{center}
    \vspace{-0.1cm}
    \begin{center}
      \caption{Minimum $L$ and its corresponding $R$ \\for $D = 0.1500$}
      \label{tbl:3}
      \vspace{-0.3cm}
      \scalebox{0.92}{
      \begin{tabular}{|c|c|c|} \hline
    Cases & Leakage $L$ & Coding rate $R$\\ \hline \hline
    \multicolumn{1}{|l|}{case (ii)} & 0.011748 & 0.270436 \\
    \multicolumn{1}{|l|}{case (iii)} & 0.002032 & 0.294424 \\
    \multicolumn{1}{|l|}{case (i)} & 0.000000 & 0.382211 \\ \hline
\end{tabular}}
    \end{center}
\end{table}

In the first example, we calculated the $L$-$D$ graph of theoretical limits in case (i) $\mathcal{E} = \mathcal{K}$, case (ii) $\mathcal{E} = \mathcal{R}$, and case (iii) $\mathcal{R} \subset \mathcal{E} \subset \mathcal{K}$ (Fig. \ref{fig:2}). As a result, the achievable privacy leakage $L$ becomes small as $D$ becomes large. The second example calculated the $R$-$D$ graph of theoretical limits in cases (i), (ii), and (iii) (Fig. \ref{fig:3}). We can see that the minimum coding rates for given $D$ coincide in all cases if we do not impose any restrictions on  the value of $L$. In the third example, we calculated the optimal privacy leakage $L$ for fixed $D$ and the corresponding coding rates $R$ in cases (i), (ii), and (iii) (Tables \ref{tbl:1}--\ref{tbl:3}). As a result, the optimal privacy leakage in cases (i) and (iii) is smaller than the one in case (ii) whereas for the optimal privacy leakage, the achievable coding rates in cases (i) and (iii) is larger than the one in case (ii).

Next, we discuss these results. In Fig. \ref{fig:2}, in comparison with each case, we can verify that for given $D$ the more private information is encoded, the smaller the achievable minimum privacy leakage is. Fig. \ref{fig:3} suggests that if the coding rate should be minimized, it suffices to encode only the public attributes. This result is evident from our main theorem (Theorem \ref{thm:2}) because the condition on the choice of test channel $p(\hat{x}_{\mathcal{R}} | x_{\mathcal{E}})$ in case (i) is weaker than the one in case (ii), and if an appropriate test channel is taken in case (i), it is also appropriate in case (ii). It is indicated that the achievable region in case (ii) is also the one in cases (i) and (iii). The opposite is not the case. From Tables \ref{tbl:1}--\ref{tbl:3}, we can confirm the trade-off between the optimal privacy leakage $L$ for fixed $D$ and the corresponding coding rate $R$ in comparison with each case.

Some readers may suspect that case (i) is the best encoded information because the achievable region in cases (ii) and (iii) is the one in case (i). However, this is not the case. As mentioned in Sect. \ref{sec:1}, in the practical aspect, the information givers are willing to protect private information as much as possible to not only a third party but also aggregators (encoder). In case (i), all private information is revealed to the encoder. In that sense, case (i) is not necessarily the best case. In contrast, In case (ii), evidently it is not the best case because of the expression of the achievable region (\eqref{eq:3.6}).

To summarize these discussion, neither cases (i) nor (ii) are necessarily the best cases. It is crucial to select the encoded information adequately in case (iii). This is the significance of the encoded set $\mathcal{E}$ to design database in view of privacy-utility trade-off.

\section{Conclusion}

In this paper, for the privacy-constrained source coding, we characterized the achievable region $\mathcal{C}(\mathcal{E})$ in the case where public information and a part of private information are encoded. The derived characterization in  this paper reduces to the results given in \cite{a} when the encoded set $\mathcal{E}$ is either $\mathcal{K}$ or $\mathcal{R}$, and thus the given expression is ``unified". In addition, we confirmed that the achievable region is a convex set. The theoretical limits may be calculated efficiently by, for example, the Arimoto-Blahut algorithm \cite{c}, \cite{d}. Moreover, numerical examples indicate that neither $\mathcal{E} = \mathcal{K}$ nor $\mathcal{E} = \mathcal{R}$ are necessarily the best cases as the encoded alphabet and it is important to choose appropriate $\mathcal{E}$ according to the requirement of the system.

As future work, strong converse for the coding system investigated in this study is an interesting research topic. Afterward, finite-length analysis \cite{q} will be a next  challenging topic.





\appendices
\section{Proof of the Markov Chain $X_{\mathcal{E}^{\mathrm{c}}}$--$X_{\mathcal{E}}$--$\hat{X}_{\mathcal{R}}$ in Converse Part}\label{apt:A}
~~Letting $p_{i}(x_{\mathcal{E}, i}, x_{\mathcal{E}^{\mathrm{c}}, i}, \hat{x}_{\mathcal{R}, i})$ be the conditional distribution given $Q = i$,
\begin{align}
p_{i}(x_{\mathcal{E}, i}, x_{\mathcal{E}^{\mathrm{c}}, i}, \hat{x}_{\mathcal{R}, i})
&=\sum_{\substack{x_{\mathcal{E}, k}:\\k\neq i}}\sum_{\substack{x_{\mathcal{E}^{\mathrm{c}}, k}:\\k\neq i}}\sum_{\substack{\hat{x}_{\mathcal{R}, k}:\\k\neq i}}p(x_{\mathcal{E}}^{n}, x_{\mathcal{E}^{\mathrm{c}}}^{n}, \hat{x}_{\mathcal{R}}^{n})\nonumber\\
&=\sum_{\substack{x_{\mathcal{E}, k}:\\k\neq i}}\sum_{\substack{x_{\mathcal{E}^{\mathrm{c}}, k}:\\k\neq i}}p(x_{\mathcal{E}}^{n}, x_{\mathcal{E}^{\mathrm{c}}}^{n}, \hat{x}_{\mathcal{R}, i})\nonumber\\
&\stackrel{\rm{(a)}}{=}\sum_{\substack{x_{\mathcal{E}, k}:\\k\neq i}}\sum_{\substack{x_{\mathcal{E}^{\mathrm{c}}, k}:\\k\neq i}}p_{i}(x_{\mathcal{E}}^{n}, \hat{x}_{\mathcal{R}, i})p(x_{\mathcal{E}^{\mathrm{c}}}^{n}|x_{\mathcal{E}}^{n})\nonumber\\
&=\sum_{\substack{x_{\mathcal{E}, k}:\\k\neq i}}p_{i}(x_{\mathcal{E}}^{n}, \hat{x}_{\mathcal{R}, i})\sum_{\substack{x_{\mathcal{E}^{\mathrm{c}}, k}:\\k\neq i}}p(x_{\mathcal{E}^{\mathrm{c}}}^{n}|x_{\mathcal{E}}^{n})\nonumber\\
&\stackrel{\rm{(b)}}{=}\sum_{\substack{x_{\mathcal{E}, k}:\\k\neq i}}p_{i}(x_{\mathcal{E}}^{n}, \hat{x}_{\mathcal{R}, i})\cdot\nonumber\\
&\hspace{4.5mm}\sum_{\substack{x_{\mathcal{E}^{\mathrm{c}}, k}:\\k\neq i}}\left(\prod_{l=1}^{n}p(x_{\mathcal{E}^{\mathrm{c}}, l}|x_{\mathcal{E}, l})\right)\nonumber\\
&=p_{i}(x_{\mathcal{E}, i},\hat{x}_{\mathcal{R}, i})p(x_{\mathcal{E}^{\mathrm{c}}, i}|x_{\mathcal{E}, i})\nonumber\\
&=p(x_{\mathcal{E}, i})p(x_{\mathcal{E}^{\mathrm{c}}, i}|x_{\mathcal{E}, i})p_{i}(\hat{x}_{\mathcal{R}, i}|x_{\mathcal{E}, i})\nonumber\\
&=p(x_{\mathcal{E}, i},x_{\mathcal{E}^{\mathrm{c}}, i})p_{i}(\hat{x}_{\mathcal{R}, i}|x_{\mathcal{E}, i}), 
\end{align}
where
\begin{itemize}
\item[(a)] is due to the Markov chain $X_{\mathcal{E}^{\mathrm{c}}}^{n}$--$X_{\mathcal{E}}^{n}$--$\hat{X}_{\mathcal{R}, i}$,
\item[(b)] follows from the stationary memoryless source.
\end{itemize}
Therefore, we can obtain the Markov chain $X_{\mathcal{E}^{\mathrm{c}}, i}$--$X_{\mathcal{E}, i}$--$\hat{X}_{\mathcal{R}, i}$. For the marginal distribution, we can show that
\begin{align}
p(x_{\mathcal{E}},x_{\mathcal{E}^{\mathrm{c}}},\hat{x}_{\mathcal{R}})
&\stackrel{\rm{(c)}}{=}\frac{1}{n}\sum_{i=1}^{n}{p_{i}(x_{\mathcal{E}}, x_{\mathcal{E}^{\mathrm{c}}}, \hat{x}_{\mathcal{R}})} \nonumber\\
&\stackrel{\rm{(d)}}{=}\frac{1}{n}\sum_{i=1}^{n}{p_{i}(x_{\mathcal{E}},x_{\mathcal{E}^{\mathrm{c}}})p_{i}(\hat{x}_{\mathcal{R}}|x_{\mathcal{E}})}\nonumber\\
&\stackrel{\rm{(e)}}{=}p(x_{\mathcal{E}}, x_{\mathcal{E}^{\mathrm{c}}})\cdot\frac{1}{n}\sum_{i=1}^{n}{p_{i}(\hat{x}_{\mathcal{R}}|x_{\mathcal{E}})}\nonumber\\
&\stackrel{\rm{(f)}}{=}p(x_{\mathcal{E}}, x_{\mathcal{E}^{\mathrm{c}}})p(\hat{x}_{\mathcal{R}}|x_{\mathcal{E}}), 
\end{align}
where
\begin{itemize}
\item[(c)] follows because
\begin{align}
p(x_{\mathcal{E}}, x_{\mathcal{E}^{\mathrm{c}}}, \hat{x}_{\mathcal{R}})
&=\sum_{i=1}^{n}\Pr\{Q=i\}{p_{i}(x_{\mathcal{E}}, x_{\mathcal{E}^{\mathrm{c}}}, \hat{x}_{\mathcal{R}})}, 
\end{align}
\item[(d)] is due to the Markov chain $X_{\mathcal{E}^{\mathrm{c}}, i}$--$X_{\mathcal{E}, i}$--$\hat{X}_{\mathcal{R}, i}$,
\item[(e)] follows from the stationary memoryless source,
\item[(f)] follows because
\begin{align}
p(\hat{x}_{\mathcal{R}}|x_{\mathcal{E}})&=\sum_{i=1}^{n}\Pr\{Q=i\}{p_{i}(\hat{x}_{\mathcal{R}}|x_{\mathcal{E}})}.
\end{align}
\end{itemize}
Therefore, we can obtain the Markov chain $X_{\mathcal{E}^{\mathrm{c}}}$--$X_{\mathcal{E}}$--$\hat{X}_{\mathcal{R}}$. We complete the proof. \hspace{\fill}$\Box$

\section{Proof of Equation \eqref{eq:3.50}}\label{apx:B}
~~From $\tilde{\mathcal{A}}(j)\subseteq\mathcal{B}(j)$ for $j=1, 2, \ldots, M_{n}-1$,
\begin{align}
\Pr\{X_{\mathcal{K}}^{n} \in \mathcal{B}(j)\} = \Pr\{X_{\mathcal{K}}^{n} \in \tilde{\mathcal{A}}(j)\} + \Pr\{X_{\mathcal{K}}^{n} \in \mathcal{B}(j)\setminus\tilde{\mathcal{A}}(j)\}.\label{eq:B.1}
\end{align}
If $x_{\mathcal{K}}^{n} \in \mathcal{B}(j)\setminus\tilde{\mathcal{A}}(j)$, then $x_{\mathcal{E}}^{n} \in T_{\delta}^{n}(X_{\mathcal{E}}|\hat{x}_{\mathcal{R}}^{n}(j))$ and $(x_{\mathcal{E}}^{n}, x_{\mathcal{E}^{\mathrm{c}}}^{n}) \notin T_{2\delta}^{n}(X_{\mathcal{K}} | \hat{x}_{\mathcal{R}}^{n}(j))$, and thus we have $x_{\mathcal{E}^{\mathrm{c}}}^{n} \notin T_{\delta}^{n}(X_{\mathcal{E}^{\mathrm{c}}} | x_{\mathcal{E}}^{n}, \hat{x}_{\mathcal{R}}^{n}(j))$ from Lemma \ref{lem:4}. Then,
\begin{align}
x_{\mathcal{K}}^{n} \in \mathcal{B}(j)\setminus\tilde{\mathcal{A}}(j)\Longrightarrow
&x_{\mathcal{E}}^{n} \in T_{\delta}^{n}(X_{\mathcal{E}} | \hat{x}_{\mathcal{R}}^{n}(j)), \nonumber\\
&x_{\mathcal{E}^{\mathrm{c}}}^{n} \notin T_{\delta}^{n}(X_{\mathcal{E}^{\mathrm{c}}} | x_{\mathcal{E}}^{n}, \hat{x}_{\mathcal{R}}^{n}(j))
\end{align}
We can prove that
\begin{align}
&\Pr\{X_{\mathcal{K}}^{n} \in \mathcal{B}(j)\setminus\tilde{\mathcal{A}}(j)\}\nonumber\\
&\le \Pr\{X_{\mathcal{E}}^{n} \in T_{\delta}^{n}(X_{\mathcal{E}} | \hat{x}_{\mathcal{R}}^{n}(j)), X_{\mathcal{E}^{\mathrm{c}}}^{n} \notin T_{\delta}^{n}(X_{\mathcal{E}^{\mathrm{c}}} | X_{\mathcal{E}}^{n}, \hat{x}_{\mathcal{R}}^{n}(j))\}\nonumber\\
&=\sum_{x_{\mathcal{E}}^{n} \in T_{\delta}^{n}(X_{\mathcal{E}} | \hat{x}_{\mathcal{R}}^{n}(j))}\Pr\{X_{\mathcal{E}}^{n}=x_{\mathcal{E}}^{n}\}\cdot\nonumber\\
&\hspace{4.5mm}\Pr\{X_{\mathcal{E}^{\mathrm{c}}}^{n} \notin T_{\delta}^{n}(X_{\mathcal{E}^{\mathrm{c}}} | {x_{\mathcal{E}}^{n}}, \hat{x}_{\mathcal{R}}^{n}(j)) | X_{\mathcal{E}}^{n} = x_{\mathcal{E}}^{n}\}\nonumber\\
&\stackrel{\rm{(a)}}{=}\sum_{x_{\mathcal{E}}^{n} \in T_{\delta}^{n}(X_{\mathcal{E}} | \hat{x}_{\mathcal{R}}^{n}(j))}\Pr\{X_{\mathcal{E}}^{n}=x_{\mathcal{E}}^{n}\}\cdot\nonumber\\
&\hspace{4.5mm}\Pr\{X_{\mathcal{E}^{\mathrm{c}}}^{n} \notin T_{\delta}^{n}(X_{\mathcal{E}^{\mathrm{c}}} | {x_{\mathcal{E}}^{n}}, \hat{x}_{\mathcal{R}}^{n}(j)) | X_{\mathcal{E}}^{n} = x_{\mathcal{E}}^{n}, \hat{X}_{\mathcal{R}}^{n} = \hat{x}_{\mathcal{R}}^{n}(j)\}\nonumber\\
&\stackrel{\rm{(b)}}{\le}\sum_{x_{\mathcal{E}}^{n} \in T_{\delta}^{n}(X_{\mathcal{E}} | \hat{x}_{\mathcal{R}}^{n}(j))}\Pr\{X_{\mathcal{E}}^{n}=x_{\mathcal{E}}^{n}\}\cdot2|\mathcal{X}_{\mathcal{E}^{\mathrm{c}}}|\cdot|\mathcal{X}_{\mathcal{E}}|\cdot|\hat{\mathcal{X}}_{\mathcal{R}}|\mathrm{e}^{-2\delta^{2}n}\nonumber\\
&\le 2|\mathcal{X}_{\mathcal{K}}|\cdot|\hat{\mathcal{X}}_{\mathcal{R}}|\mathrm{e}^{-2\delta^{2}n}, \label{eq:B.7}
\end{align}
where
\begin{itemize}
\item[(a)] is due to the Markov chain $X_{\mathcal{E}^{\mathrm{c}}}^{n}$--$X_{\mathcal{E}}^{n}$--$\hat{X}_{\mathcal{R}}^{n}$,
\item[(b)] follows from Lemma \ref{lem:5}.
\end{itemize}
From equations \eqref{eq:B.1} and \eqref{eq:B.7}, we can obtain
\begin{align}
|\Pr\{X_{\mathcal{K}}^{n} \in \mathcal{B}(j)\} - \Pr\{X_{\mathcal{K}}^{n} \in \tilde{\mathcal{A}}(j)\}| \le 2|\mathcal{X}_{\mathcal{K}}|\cdot|\hat{\mathcal{X}}_{\mathcal{R}}|\mathrm{e}^{-2\delta^{2}n}.\label{eq:B.8}
\end{align}
We complete the proof of \eqref{eq:3.50}. \hspace{\fill}$\Box$

\section{Proof of Existence of Code Satisfying \eqref{eq:4.25}--\eqref{eq:4.29}}\label{apx:C}
We first set $M_{n} \coloneqq 2^{nR}$ and $r_{n} \coloneqq \frac{1}{n}\log{M_{n}}$. Then, we obviously have \eqref{eq:4.25}.

From the union upper bound,
\begin{align}
&\Pr\left\{X_{\mathcal{E}}^{n} \notin \bigcup_{j=1}^{M_{n}-1}\mathcal{A}(j)\right\} \nonumber\\
&\le \Pr\{X_{\mathcal{E}}^{n} \notin T_{\delta}^{n}(X_{\mathcal{E}})\}\nonumber\\
&\hspace{4.5mm}+\Pr\{X_{\mathcal{E}}^{n} \in T_{\delta}^{n}(X_{\mathcal{E}}), X_{\mathcal{E}}^{n} \notin T_{\delta}^{n}(X_{\mathcal{E}}|\hat{x}_{\mathcal{R}}^{n}(j)) \nonumber\\
&\hspace{12.9mm}\mathrm{\ for\ all}\ j=1, {2}, \ldots,  M_{n}-1\}.\label{eq:D.1}
\end{align}
From Lemma \ref{lem:5}, the first term in \eqref{eq:D.1} is bounded as
\begin{align}
\Pr\{X_{\mathcal{E}}^{n} \notin T_{\delta}^{n}(X_{\mathcal{E}})\} \le 2|\mathcal{X}_{\mathcal{E}}|\mathrm{e}^{-2\delta^{2}n}.\label{eq:D.2}
\end{align}
We consider the expectation of the second term in \eqref{eq:D.1} by random coding. Hereafter, we denote the random variable corresponding to the reproduced sequence $\hat{x}_{\mathcal{R}}^{n}(j)$ as $\hat{X}_{\mathcal{R}}^{n}(j)$. For notational simplicity, we use the abbreviation 
\begin{align}
&\Pr\{X_{\mathcal{E}}^{n} \notin T_{\delta}^{n}(X_{\mathcal{E}}|\hat{X}_{\mathcal{R}}^{n}(j))\nonumber\\
&\hspace{6mm}\mathrm{for\ all}\ j=1, 2, \ldots,  M_{n}-1 | X_{\mathcal{E}}^{n} = x_{\mathcal{E}}^{n}\}\nonumber\\
&= \Pr\{{x}_{\mathcal{E}}^{n} \notin T_{\delta}^{n}(X_{\mathcal{E}}|\hat{X}_{\mathcal{R}}^{n}(j))\nonumber\\
&\hspace{10.2mm}\mathrm{for\ all}\ j=1, 2, \ldots,  M_{n}-1\}, \label{eq:77}
\end{align}
and then
\begin{align}
&\mathbb{E}[\Pr\{X_{\mathcal{E}}^{n} \in T_{\delta}^{n}(X_{\mathcal{E}}), X_{\mathcal{E}}^{n} \notin T_{\delta}^{n}(X_{\mathcal{E}}|\hat{X}_{\mathcal{R}}^{n}(j))\nonumber\\
&\hspace{9mm}\mathrm{for\ all}\ j=1, {2}, \ldots,  M_{n}-1\}]\nonumber\\
&=\sum_{x_{\mathcal{E}}^{n} \in T_{\delta}^{n}(X_{\mathcal{E}})} p(x_{\mathcal{E}}^{n})\mathbb{E}\Big{[}\Pr\{X_{\mathcal{E}}^{n} \notin T_{\delta}^{n}(X_{\mathcal{E}}|\hat{X}_{\mathcal{R}}^{n}(j))\nonumber\\
&\hspace{9mm}\mathrm{for\ all}\ j=1, 2, \ldots,  M_{n}-1 | X_{\mathcal{E}}^{n} = x_{\mathcal{E}}^{n}\}\Big{]}\nonumber\\
&\stackrel{\rm{(a)}}{=}\sum_{x_{\mathcal{E}}^{n} \in T_{\delta}^{n}(X_{\mathcal{E}})} p(x_{\mathcal{E}}^{n})\mathbb{E}\Big{[}\Pr\{{x}_{\mathcal{E}}^{n} \notin T_{\delta}^{n}(X_{\mathcal{E}}|\hat{X}_{\mathcal{R}}^{n}(j))\nonumber\\
&\hspace{9.6mm}\mathrm{for\ all}\ j=1, 2, \ldots,  M_{n}-1\}\Big{]}\nonumber\\
&=\sum_{x_{\mathcal{E}}^{n} \in T_{\delta}^{n}(X_{\mathcal{E}})} p(x_{\mathcal{E}}^{n})\left(\prod_{j=1}^{M_{n}-1}\mathbb{E}\left[\Pr\{{x}_{\mathcal{E}}^{n} \notin T_{\delta}^{n}(X_{\mathcal{E}}|\hat{X}_{\mathcal{R}}^{n}(j))\}\right]\right)\nonumber\\
&\stackrel{\rm{(b)}}{=}\sum_{x_{\mathcal{E}}^{n} \in T_{\delta}^{n}(X_{\mathcal{E}})} p(x_{\mathcal{E}}^{n})\left(\mathbb{E}\left[\Pr\{{x}_{\mathcal{E}}^{n} \notin T_{\delta}^{n}(X_{\mathcal{E}}|\hat{X}_{\mathcal{R}}^{n}(1))\}\right]\right)^{M_{n}-1} \nonumber\\
&\stackrel{\rm{(c)}}{\le} \exp\left\{-2^{n(R-I(X_{\mathcal{E}};\hat{X}_{\mathcal{R}})-\frac{1}{n}-\tau)}\right\}\nonumber\\
&\stackrel{\rm{(d)}}{\le} \exp\left\{-2^{2\delta^{2}n}\right\}, \label{eq:C.10}
\end{align}
where
\begin{itemize}
\item[(a)] is owing to \eqref{eq:77},
\item[(b)] is due to the symmetry about indexes of random coding,
\item[(c)] follows from the same way as in \cite[Sect. 3.6.3]{l},
\item[(d)] because $\delta$ is fixed to satisfy \eqref{eq:4.17}.
\end{itemize}
From \eqref{eq:D.2} and \eqref{eq:C.10}, we obtain
\begin{align}
\mathbb{E}\left[\Pr\left\{X_{\mathcal{E}}^{n} \notin \bigcup_{j=1}^{M_{n}-1}\mathcal{A}(j)\right\}\right] \le (2|\mathcal{X}_{\mathcal{E}}| + 1)\mathrm{e}^{-2\delta^{2}n}. \label{eq:C.23}
\end{align}
Therefore, there exists at least one codebook satisfying \eqref{eq:4.27} in the ensembles obtained by random coding.

Hereafter, codebook $\mathcal{C}$ is fixed to satisfy \eqref{eq:4.27}. That is, codebook $\mathcal{C}$ satisfies
\begin{align}
\Pr\left\{X_{\mathcal{E}}^{n} \notin \bigcup_{j=1}^{M_{n}-1}\mathcal{A}(j)\right\} \le (2|\mathcal{X}_{\mathcal{E}}| + 1)\mathrm{e}^{-2\delta^{2}n}. \label{eq:C.6}
\end{align}
We evaluate the distortion function for each $j$.
\begin{itemize}
\item[(i)] $j=1, 2, \ldots, M_{n}-1$:
\begin{align}
&d(x_{\mathcal{R}}^{n}, \hat{x}_{\mathcal{R}}^{n}(j))\nonumber\\
&=\frac{1}{n}\sum_{a\in \mathcal{X}_{\mathcal{R}}}\sum_{b\in \hat{\mathcal{X}}_{\mathcal{R}}}N(a, b|x_{\mathcal{R}}^{n}, \hat{x}_{\mathcal{R}}^{n}(j))d(a, b)\nonumber\\
&\stackrel{\rm{(e)}}{\le}\sum_{a\in \mathcal{X}_{\mathcal{R}}}\sum_{b\in \hat{\mathcal{X}}_{\mathcal{R}}}P_{X_{\mathcal{R}}, \hat{X}_{\mathcal{R}}}(a, b)d(a, b)\nonumber\\
&\hspace{4.5mm}+(\delta + \delta_{1}) |\mathcal{X}_{\mathcal{R}}| \cdot |\hat{\mathcal{X}}_{\mathcal{R}}|D_{\max}\nonumber\\
&=\mathbb{E}[d(X_{\mathcal{R}}, \hat{X}_{\mathcal{R}})]+(\delta + \delta_{1}) |\mathcal{X}_{\mathcal{R}}| \cdot |\hat{\mathcal{X}}_{\mathcal{R}}|D_{\max}, \label{eq:114}
\end{align}
where
\begin{itemize}
\item[(e)] because from Lemma \ref{lem:3}, if $x_{\mathcal{E}}^{n} \in T_{\delta}^{n}(X_{\mathcal{E}} | \hat{x}_{\mathcal{R}}^{n}(j))$, then $x_{\mathcal{R}}^{n} \in T_{\delta_{1}}^{n}(X_{\mathcal{R}} | \hat{x}_{\mathcal{R}}^{n}(j))$ and from Lemma \ref{lem:2}, if $\hat{x}_{\mathcal{R}}^{n}(j)\in T_{\delta}^{n}(\hat{X}_{\mathcal{R}})$ and $x_{\mathcal{R}}^{n} \in T_{\delta_{1}}^{n}(X_{\mathcal{R}} | \hat{x}_{\mathcal{R}}^{n}(j))$, then $(x_{\mathcal{R}}^{n}, \hat{x}_{\mathcal{R}}^{n}(j))\in T_{\delta + \delta_{1}}^{n}(X_{\mathcal{R}}, \hat{X}_{\mathcal{R}})$.
\end{itemize}
\item[(ii)] $j=M_{n}$:
\begin{align}
d(x_{\mathcal{R}}^{n}, \hat{x}_{\mathcal{R}}^{n}(M_{n}))
&=\frac{1}{n}\sum_{i=1}^{n}d(x_{\mathcal{R}, i}, \hat{x}_{\mathcal{R}, i})\nonumber\\
&\stackrel{\rm{(f)}}{\le} D_{\max}, \label{eq:C.5}
\end{align}
where
\begin{itemize}
\item[(f)] is due to the definition of $D_{\max} \coloneqq \underset{a\in\mathcal{X}_{\mathcal{R}}, b\in\hat{\mathcal{X}}_{\mathcal{R}}}{\max}d(a, b)$.
\end{itemize}
\end{itemize}
We consider $\Pr\{J_{n} = M_{n}\}$. From \eqref{eq:C.6},
\begin{align}
\Pr\{J_{n} = M_{n}\} 
&= \Pr\{X_{\mathcal{E}}^{n} \in \mathcal{A}(M_{n})\}\nonumber\\
&=\Pr\left\{X_{\mathcal{E}}^{n} \notin \bigcup_{j=1}^{M_{n}-1}\mathcal{A}(j)\right\}\nonumber\\
&\le (2|\mathcal{X}_{\mathcal{E}}| + 1)\mathrm{e}^{-2\delta^{2}n}.
\end{align}
Therefore, we can confirm
\begin{align}
\lim_{n \rightarrow \infty}\Pr\{J_{n} = M_{n}\} = 0.\label{eq:C.21}
\end{align}
From (i) and (ii), we can evaluate utility $u_{n}$ as below.
\begin{align}
u_{n}&\coloneqq\mathbb{E}\left[d(X_{\mathcal{R}}^{n}, \hat{X}_{\mathcal{R}}^{n})\right]\nonumber\\
&\leq\sum_{j=1}^{M_{n}-1}\Pr\{J_{n}=j\}\cdot\Big{(}\mathbb{E}[d(X_{\mathcal{R}}, \hat{X}_{\mathcal{R}})]\nonumber\\
&\hspace{4.5mm}+(\delta + \delta_{1}) |\mathcal{X}_{\mathcal{R}}| \cdot |\hat{\mathcal{X}}_{\mathcal{R}}|D_{\max}\Big{)}+\Pr\{J_{n}=M_{n}\}\cdot D_{\max}\nonumber\\
&\stackrel{\rm{(g)}}{\le}\mathbb{E}[d(X_{\mathcal{R}}, \hat{X}_{\mathcal{R}})]+(\delta + \delta_{1}) |\mathcal{X}_{\mathcal{R}}| \cdot |\hat{\mathcal{X}}_{\mathcal{R}}|D_{\max}+\tau
\end{align}
for all sufficiently large $n$, where
\begin{itemize}
\item[(g)] follows from \eqref{eq:C.21}.
\end{itemize}
Thus, we obtain \eqref{eq:4.26}.

Next, we show that the probability that random vector $X_{\mathcal{K}}^{n}$ are not included in the set $\bigcup_{j=1}^{M_{n}-1}\tilde{\mathcal{A}}(j)$, is sufficiently small. First, notice that
\begin{align}
x_{\mathcal{K}}^{n} \notin \bigcup_{j=1}^{M_{n}-1}\tilde{\mathcal{A}}(j)
\Longrightarrow 
&x_{\mathcal{E}}^{n} \notin \bigcup_{j=1}^{M_{n}-1}\mathcal{A}(j)\nonumber\\
&\hspace{7mm}\mathrm{or}\nonumber\\
&x_{\mathcal{E}}^{n} \in \mathcal{A}(j_{0}), \nonumber\\
&(x_{\mathcal{E}}^{n}, x_{\mathcal{E}^{\mathrm{c}}}^{n}) \notin  T_{2\delta}^{n}(X_{\mathcal{K}}|\hat{x}_{\mathcal{R}}^{n}(j_{0}))\nonumber\\
&\mathrm{for}\ j_{0} = f_{n}(x_{\mathcal{E}}^{n}),
\end{align}
where $j_{0}$ is the index such that $f_{n}(x_{\mathcal{E}}^{n}) = j_{0}$ for $1 \le j_{0} \le M_{n} -1$. Therefore, by the union upper bound,
\begin{align}
&\Pr\left\{X_{\mathcal{K}}^{n} \notin \bigcup_{j=1}^{M_{n}-1}\tilde{\mathcal{A}}(j)\right\}\nonumber\\
&\le \Pr\left\{X_{\mathcal{E}}^{n} \notin \bigcup_{j=1}^{M_{n}-1}\mathcal{A}(j)\right\}\nonumber\\
&\hspace{4.5mm}+\Pr\{X_{\mathcal{E}}^{n} \in \mathcal{A}(j_{0}), (X_{\mathcal{E}}^{n}, X_{\mathcal{E}^{\mathrm{c}}}^{n}) \notin T_{2\delta}^{n}(X_{\mathcal{K}}|\hat{x}_{\mathcal{R}}^{n}(j_{0}))\nonumber\\
&\hspace{14mm}\mathrm{for}\ j_{0} = f_{n}(X_{\mathcal{E}}^{n})\}.\label{eq:C.14}
\end{align}
We evaluate each term in \eqref{eq:C.14}.
\begin{itemize}
\item[(i)] The first term:
\begin{align}
\Pr\left\{X_{\mathcal{E}}^{n} \notin \bigcup_{j=1}^{M_{n}-1}\mathcal{A}(j)\right\}
&\stackrel{\rm{(h)}}{\le} (2|\mathcal{X}_{\mathcal{E}}| + 1)\mathrm{e}^{-2\delta^{2}n}, \label{eq:C.16}
\end{align}
where
\begin{itemize}
\item[(h)] is because of \eqref{eq:C.6}.
\end{itemize}
\item[(ii)] The second term:\\
~If the event in the second term occurs, $x_{\mathcal{E}}^{n} \in T_{\delta}^{n} (X_{\mathcal{E}} | \hat{x}_{\mathcal{R}}^{n}(j_{0}))$ and $(x_{\mathcal{E}}^{n}, x_{\mathcal{E}^{\mathrm{c}}}^{n}) \notin T_{2\delta}^{n}(X_{\mathcal{K}}|\hat{x}_{\mathcal{R}}^{n}(j_{0}))$. Therefore, from Lemma \ref{lem:4}, $x_{\mathcal{E}^{\mathrm{c}}}^{n} \notin T_{\delta}^{n}(X_{{\mathcal{E}^{\mathrm{c}}}} | x_{\mathcal{E}}^{n}, \hat{x}_{\mathcal{R}}^{n}(j_{0}))$ holds. Hence,
\begin{align}
&\Pr\{X_{\mathcal{E}}^{n} \in \mathcal{A}(j_{0}), (X_{\mathcal{E}}^{n}, X_{\mathcal{E}^{\mathrm{c}}}^{n}) \notin T_{2\delta}^{n}(X_{\mathcal{K}}|\hat{x}_{\mathcal{R}}^{n}(j_{0}))\nonumber\\
&\hspace{6mm}\mathrm{for}\ j_{0} = f_{n}(X_{\mathcal{E}}^{n})\}\nonumber\\
&\le \Pr\{X_{\mathcal{E}}^{n} \in \mathcal{A}(j_{0}), X_{\mathcal{E}^{\mathrm{c}}}^{n} \notin T_{\delta}^{n}(X_{\mathcal{E}^{\mathrm{c}}} | X_{\mathcal{E}}^{n}, \hat{x}_{\mathcal{R}}^{n}(j_{0}))\}\nonumber\\
&\le\sum_{j = 1}^{M_{n}-1}\sum_{x_{\mathcal{E}}^{n} \in \mathcal{A}(j)}\Pr\{X_{\mathcal{E}}^{n} = x_{\mathcal{E}}^{n}\}\cdot\nonumber\\&\hspace{4.5mm}\Pr\{X_{\mathcal{E}^{\mathrm{c}}}^{n} \notin T_{\delta}^{n}(X_{\mathcal{E}^{\mathrm{c}}} | x_{\mathcal{E}}^{n}, \hat{x}_{\mathcal{R}}^{n}(j)) | X_{\mathcal{E}}^{n} = x_{\mathcal{E}}^{n}\}\nonumber\\
&\stackrel{\rm{(i)}}{=}\sum_{j = 1}^{M_{n}-1}\sum_{x_{\mathcal{E}}^{n} \in \mathcal{A}(j)}\Pr\{X_{\mathcal{E}}^{n} = x_{\mathcal{E}}^{n}\}\cdot\nonumber\\
&\hspace{4.5mm}\Pr\{X_{\mathcal{E}^{\mathrm{c}}}^{n} \notin T_{\delta}^{n}(X_{\mathcal{E}^{\mathrm{c}}} | x_{\mathcal{E}}^{n}, \hat{x}_{\mathcal{R}}^{n}(j)) | X_{\mathcal{E}}^{n} = x_{\mathcal{E}}^{n}, \hat{X}_{\mathcal{R}}^{n} = \hat{x}_{\mathcal{R}}^{n}(j)\}\nonumber\\
&\stackrel{\rm{(j)}}{\le} \sum_{j = 1}^{M_{n}-1}\sum_{x_{\mathcal{E}}^{n} \in \mathcal{A}(j)}\Pr\{X_{\mathcal{E}}^{n} = x_{\mathcal{E}}^{n}\} \cdot 2|\mathcal{X}_{\mathcal{E}^{\mathrm{c}}}| \cdot |\mathcal{X}_{\mathcal{E}}| \cdot |\hat{\mathcal{X}}_{\mathcal{R}}| \mathrm{e}^{-2\delta^{2}n}\nonumber\\
&\stackrel{\rm{(k)}}{\le} 2|\mathcal{X}_{\mathcal{K}}| \cdot |\hat{\mathcal{X}}_{\mathcal{R}}| \mathrm{e}^{-2\delta^{2}n}, \label{eq:C.27}
\end{align}
where
\begin{itemize}
\item[(i)] is due to the Markov chain $X_{\mathcal{E}^{\mathrm{c}}}^{n}-X_{\mathcal{E}}^{n}-\hat{X}_{\mathcal{R}}^{n}$,
\item[(j)] follows since $x_{\mathcal{E}}^{n} \in T_{\delta}^{n}(X_{\mathcal{E}} | \hat{x}_{\mathcal{R}}^{n}(j_{0}))$ and Lemma \ref{lem:5},
\item[(k)] follows because $\mathcal{A}(j)$ are disjoint for each $j$.
\end{itemize}
\end{itemize}
From \eqref{eq:C.14}, \eqref{eq:C.16}, and \eqref{eq:C.27},
\begin{align}
\Pr\left\{X_{\mathcal{K}}^{n} \notin \bigcup_{j=1}^{M_{n}-1}\tilde{\mathcal{A}}(j)\right\} \le 4|\mathcal{X}_{{\mathcal{K}}}| \cdot |\hat{\mathcal{X}}_{\mathcal{R}}|\mathrm{e}^{-2\delta^{2}n}.
\end{align}
Therefore, for sufficiently large $n$,
\begin{align}
\Pr\left\{X_{\mathcal{K}}^{n} \notin \bigcup_{j=1}^{M_{n}-1}\tilde{\mathcal{A}}(j)\right\} \le \tau,
\end{align}
and we obtain \eqref{eq:4.28}.

From Lemma \ref{lem:6}, for sufficiently large $n$ to stochastic matrix $W\!:~\hat{\mathcal{X}}_{\mathcal{R}} \rightarrow \mathcal{X}_{\mathcal{K}}$ and $\hat{x}_{\mathcal{R}}^{n}(j) \in T_{\delta}^{n}(\hat{X}_{\mathcal{R}})$ we can show that
\begin{align}
&\left|\frac{1}{n}\log|T_{\delta_{2}}^{n}(X_{\mathcal{K}} | \hat{x}_{\mathcal{R}}^{n}(j))| - H(X_{\mathcal{K}} | \hat{X}_{\mathcal{R}})\right| \le \tau, \label{eq:C.40}\\
&\delta_{2} \coloneqq \frac{\delta}{|\mathcal{X}_{\mathcal{E}^{\mathrm{c}}}|}. \nonumber
\end{align}
Also, we can show from \eqref{eq:C.40} that
\begin{align}
2^{n\{H(X_{\mathcal{K}} | \hat{X}_{\mathcal{R}})-\tau\}}
\le |T_{\delta_{2}}^{n}(X_{\mathcal{K}}| \hat{x}_{\mathcal{R}}^{n}(j))| \le 2^{n\{H(X_{\mathcal{K}} | \hat{X}_{\mathcal{R}}) + \tau\}}.\label{eq:C.31}
\end{align}
From the definition of $\tilde{\mathcal{A}}(j)$ and $T_{\delta_{2}}^{n}(X_{\mathcal{K}} | \hat{x}_{\mathcal{R}}^{n}(j))$ and Lemma \ref{lem:3}, for $j = 1, 2, \ldots, M_{n} - 1$, we have
\begin{align}
x_{\mathcal{K}}^{n} \in \tilde{A}(j) &\Longleftrightarrow\hspace{-0.8mm}
\begin{cases}
x_{\mathcal{E}}^{n} \in T_{\delta}^{n}(X_{\mathcal{E}} | \hat{x}_{\mathcal{R}}^{n}(j)) \\
x_{\mathcal{K}}^{n} \in T_{2\delta}^{n}(X_{\mathcal{K}} | \hat{x}_{\mathcal{R}}^{n}(j))
\end{cases} \\
x_{\mathcal{K}}^{n} \in T_{\delta_{2}}^{n}(X_{\mathcal{K}} | \hat{x}_{\mathcal{R}}^{n}(j)) &\Longrightarrow
\begin{cases}
x_{\mathcal{E}}^{n} \in T_{\delta}^{n}(X_{\mathcal{E}} | \hat{x}_{\mathcal{R}}^{n}(j)) \\
x_{\mathcal{K}}^{n} \in T_{2\delta}^{n}(X_{\mathcal{K}} | \hat{x}_{\mathcal{R}}^{n}(j))
\end{cases}
\end{align}
This means
\begin{align}
T_{\delta_{2}}^{n}(X_{\mathcal{K}}| \hat{x}_{\mathcal{R}}^{n}(j)) &\subseteq \tilde{\mathcal{A}}(j)\nonumber\\
\Longrightarrow|T_{\delta_{2}}^{n}(X_{\mathcal{K}}| \hat{x}_{\mathcal{R}}^{n}(j))| &\le |\tilde{\mathcal{A}}(j)|. \label{eq:C.46}
\end{align}
Therefore, from \eqref{eq:C.31} and \eqref{eq:C.46},
\begin{align}
|\tilde{\mathcal{A}}(j)| \ge 2^{n\{H(X_{\mathcal{K}} | \hat{X}_{\mathcal{R}})-\tau\}},
\end{align}
and we obtain \eqref{eq:4.29}.\hspace{\fill}$\Box$

\section{Derivation of Inequality in \eqref{eq:3.62}}\label{apx:D}
We derive the inequality in \eqref{eq:3.62}. To write notation concisely, for every $x_{\mathcal{H}}^{n} \in \mathcal{X}_{\mathcal{H}}^{n}$ and each $j = 1, 2, \ldots, M_{n}$, we define $P_{n}(j)$, $Q_{n}(j)$, $\tilde{P}_{n}(x_{\mathcal{H}}^{n}, j)$, and $\tilde{Q}_{n}(x_{\mathcal{H}}^{n}, j)$ as follows:
\begin{align}
P_{n}(j)&\coloneqq\Pr\{X_{\mathcal{K}}^{n} \in \mathcal{B}(j)\}, \\
Q_{n}(j)&\coloneqq\Pr\{X_{\mathcal{K}}^{n} \in \tilde{\mathcal{A}}(j)\}, \\
\tilde{P}_{n}(x_{\mathcal{H}}^{n}, j)&\coloneqq\Pr\{X_{\mathcal{H}}^{n} = x_{\mathcal{H}}^{n}, X_{\mathcal{K}}^{n} \in \mathcal{B}(j)\}, \\
\tilde{Q}_{n}(x_{\mathcal{H}}^{n}, j)&\coloneqq\Pr\{X_{\mathcal{H}}^{n} = x_{\mathcal{H}}^{n}, X_{\mathcal{K}}^{n} \in \tilde{\mathcal{A}}(j)\}.
\end{align}
Then, using the notation in \cite{f} we can write each entropy as
\begin{align}
H(X_{\mathcal{K}}^{n} \in \mathcal{B}(J_{n})) &= H(P_{n}),\\
H(X_{\mathcal{K}}^{n} \in \tilde{\mathcal{A}}(J_{n})) &= H(Q_{n}),\\
H(X_{\mathcal{H}}^{n}, X_{\mathcal{K}}^{n} \in \mathcal{B}(J_{n})) &= H(\tilde{P}_{n}),\\
H(X_{\mathcal{H}}^{n}, X_{\mathcal{K}}^{n} \in \tilde{\mathcal{A}}(J_{n})) &= H(\tilde{Q}_{n}).
\end{align}
The variational distance between distributions $P_{n}$ and $Q_{n}$ is
\begin{align}
d_{\mathrm{v}}(P_{n}, Q_{n}) 
&= \sum_{j=1}^{M_{n}} |P_{n}(j) - Q_{n}(j)| \nonumber\\
&= \sum_{j=1}^{M_{n}-1} |P_{n}(j) - Q_{n}(j)|\nonumber\\
&\hspace{4.5mm} + |P_{n}(M_{n}) - Q_{n}(M_{n})|. \label{eq:D.10}
\end{align}
We evaluate each term in \eqref{eq:D.10}.
\begin{itemize}
\item[(i)] The first term:
\begin{align}
&\sum_{j=1}^{M_{n}-1} |P_{n}(j) - Q_{n}(j)| \nonumber\\
&= \sum_{j=1}^{M_{n}-1} \Pr\{X_{\mathcal{K}}^{n} \in \mathcal{B}(j)\setminus\tilde{\mathcal{A}}(j)\}\nonumber\\
&\stackrel{\rm{(a)}}{=} \Pr\left\{ X_{\mathcal{K}}^{n} \in \bigcup_{j=1}^{M_{n}-1}\mathcal{B}(j)\setminus\tilde{\mathcal{A}}(j)\right\} \nonumber\\
&= \Pr\left\{X_{\mathcal{K}}^{n} \in \bigcup_{j = 1}^{M_{n} - 1} \mathcal{B}(j)\right\} 
- \Pr\left\{X_{\mathcal{K}}^{n} \in \bigcup_{j = 1}^{M_{n} - 1} \tilde{\mathcal{A}}(j)\right\} \nonumber\\
&= \left( 1 - \Pr\left\{X_{\mathcal{K}}^{n} \in \bigcup_{j = 1}^{M_{n} - 1} \tilde{\mathcal{A}}(j)\right\}  \right) \nonumber\\
&\hspace{4.5mm}- \left( 1 - \Pr\left\{X_{\mathcal{K}}^{n} \in \bigcup_{j = 1}^{M_{n} - 1} \mathcal{B}(j)\right\}  \right) \nonumber\\
&= \Pr\left\{X_{\mathcal{K}}^{n} \notin \bigcup_{j = 1}^{M_{n} - 1} \tilde{\mathcal{A}}(j)\right\} 
- \Pr\left\{X_{\mathcal{K}}^{n} \notin \bigcup_{j = 1}^{M_{n} - 1} \mathcal{B}(j)\right\} \nonumber\\
&\le \Pr\left\{X_{\mathcal{K}}^{n} \notin \bigcup_{j = 1}^{M_{n} -1} \tilde{\mathcal{A}}(j)\right\} \nonumber\\
&\stackrel{\rm{(b)}}{\le} \tau, \label{eq:D.17}
\end{align}
where
\begin{itemize}
\item[(a)] follows because $\mathcal{B}(j) \setminus \tilde{A}(j)$ is disjoint for each $j = 1, 2, \ldots, M_{n}-1$,
\item[(b)] is owing to \eqref{eq:4.28}.
\end{itemize}
\item[(ii)] The second term:
\begin{align}
|P_{n}(M_{n}) - Q_{n}(M_{n})| 
&\stackrel{\rm{(c)}}{=} Q_{n}(M_{n}) - P_{n}(M_{n}) \nonumber\\
&\le Q_{n}(M_{n}) \nonumber\\
&= \Pr\left\{X_{\mathcal{K}}^{n} \notin \bigcup_{j = 1}^{M_{n} - 1} \tilde{\mathcal{A}}(j)\right\} \nonumber\\
&\stackrel{\rm{(d)}}{\le} \tau, \label{eq:D.21}
\end{align}
where
\begin{itemize}
\item[(c)] follows because $\mathcal{B}(M_{n}) \subseteq \tilde{\mathcal{A}}(M_{n})$,
\item[(d)] follows from \eqref{eq:4.28}.
\end{itemize}
\end{itemize}
From \eqref{eq:D.17} and \eqref{eq:D.21}, the variational distance between $P_{n}$ and $Q_{n}$ is bounded from above as
\begin{align}
d_{\mathrm{v}}(P_{n}, Q_{n}) 
&\le \tau + \tau \nonumber\\
&= 2\tau.
\end{align}

Next, the variational distance between distributions $\tilde{P}_{n}$ and $\tilde{Q}_{n}$ is
\begin{align}
d_{\mathrm{v}}(\tilde{P}_{n}, \tilde{Q}_{n})
&= \sum_{j = 1}^{M_{n}}\sum_{x_{\mathcal{H}}^{n} \in \mathcal{X}_{\mathcal{H}}^{n}} \left| \tilde{P}_{n}(x_{\mathcal{H}}^{n}, j) - \tilde{Q}_{n}(x_{\mathcal{H}}^{n}, j)  \right| \nonumber\\
&= \sum_{j = 1}^{M_{n}-1}\sum_{x_{\mathcal{H}}^{n} \in \mathcal{X}_{\mathcal{H}}^{n}} \left| \tilde{P}_{n}(x_{\mathcal{H}}^{n}, j) - \tilde{Q}_{n}(x_{\mathcal{H}}^{n}, j)  \right| \nonumber\\
&\hspace{4.5mm}+ \sum_{x_{\mathcal{H}}^{n} \in \mathcal{X}_{\mathcal{H}}^{n}} \left| \tilde{P}_{n}(x_{\mathcal{H}}^{n}, M_{n}) - \tilde{Q}_{n}(x_{\mathcal{H}}^{n}, M_{n})  \right|. \label{eq:D.24}
\end{align}
We evaluate each term in \eqref{eq:D.24}.
\begin{itemize}
\item[(i)] The first term:
\begin{align}
&\sum_{j = 1}^{M_{n}-1}\sum_{x_{\mathcal{H}}^{n} \in \mathcal{X}_{\mathcal{H}}^{n}} \left| \tilde{P}_{n}(x_{\mathcal{H}}^{n}, j) - \tilde{Q}_{n}(x_{\mathcal{H}}^{n}, j)  \right| \nonumber\\
&= \sum_{j = 1}^{M_{n}-1}\sum_{x_{\mathcal{H}}^{n} \in \mathcal{X}_{\mathcal{H}}^{n}} \Pr\{X_{\mathcal{H}}^{n} = x_{\mathcal{H}}^{n}, X_{\mathcal{K}}^{n} \in  \mathcal{B}(j)\setminus\tilde{\mathcal{A}}(j)\} \nonumber\\
&\stackrel{\rm{(e)}}{=} \sum_{x_{\mathcal{H}}^{n} \in \mathcal{X}_{\mathcal{H}}^{n}} \Pr\left\{X_{\mathcal{H}}^{n} = x_{\mathcal{H}}^{n}, X_{\mathcal{K}}^{n} \in  \bigcup_{j = 1}^{M_{n}-1}\mathcal{B}(j)\setminus\tilde{\mathcal{A}}(j)\right\} \nonumber\\
&= \Pr\left\{ X_{\mathcal{K}}^{n} \in \bigcup_{j=1}^{M_{n}-1}\mathcal{B}(j)\setminus\tilde{\mathcal{A}}(j)\right\} \nonumber\\
&= \Pr\left\{X_{\mathcal{K}}^{n} \in \bigcup_{j = 1}^{M_{n} - 1} \mathcal{B}(j)\right\} 
- \Pr\left\{X_{\mathcal{K}}^{n} \in \bigcup_{j = 1}^{M_{n} - 1} \tilde{\mathcal{A}}(j)\right\} \nonumber\\
&= \left( 1 - \Pr\left\{X_{\mathcal{K}}^{n} \in \bigcup_{j = 1}^{M_{n} - 1} \tilde{\mathcal{A}}(j)\right\}  \right) \nonumber\\
&\hspace{4.5mm}- \left( 1 - \Pr\left\{X_{\mathcal{K}}^{n} \in \bigcup_{j = 1}^{M_{n} - 1} \mathcal{B}(j)\right\}  \right) \nonumber\\
&= \Pr\left\{X_{\mathcal{K}}^{n} \notin \bigcup_{j = 1}^{M_{n} - 1} \tilde{\mathcal{A}}(j)\right\} 
- \Pr\left\{X_{\mathcal{K}}^{n} \notin \bigcup_{j = 1}^{M_{n} - 1} \mathcal{B}(j)\right\} \nonumber\\
&\le \Pr\left\{X_{\mathcal{K}}^{n} \notin \bigcup_{j = 1}^{M_{n} -1} \tilde{\mathcal{A}}(j)\right\} \nonumber\\
&\stackrel{\rm{(f)}}{\le} \tau, \label{eq:D.33}
\end{align}
where
\begin{itemize}
\item[(e)] follows since $\mathcal{B}(j)\setminus\tilde{\mathcal{A}}(j)$ is disjoint for each $j = 1, 2, \ldots, M_{n}-1$,
\item[(f)] is due to \eqref{eq:4.28}.
\end{itemize}
\item[(ii)] The second term:
\begin{align}
&\sum_{x_{\mathcal{H}}^{n} \in \mathcal{X}_{\mathcal{H}}^{n}}| \tilde{P}_{n}(x_{\mathcal{H}}^{n}, M_{n}) - \tilde{Q}_{n}(x_{\mathcal{H}}^{n}, M_{n})| \nonumber\\
&\stackrel{\rm{(g)}}{=} \sum_{x_{\mathcal{H}}^{n} \in \mathcal{X}_{\mathcal{H}}^{n}} \left(\tilde{Q}_{n}(x_{\mathcal{H}}^{n}, M_{n}) - \tilde{P}_{n}(x_{\mathcal{H}}^{n}, M_{n}) \right) \nonumber\\
&\le \sum_{x_{\mathcal{H}}^{n} \in \mathcal{X}_{\mathcal{H}}^{n}} \tilde{Q}_{n}(x_{\mathcal{H}}^{n}, M_{n}) \nonumber\\
&= Q_{n}(M_{n}) \nonumber\\
&= \Pr\left\{X_{\mathcal{K}}^{n} \notin \bigcup_{j = 1}^{M_{n} - 1} \tilde{\mathcal{A}}(j)\right\} \nonumber\\
&\stackrel{\rm{(h)}}{\le} \tau, \label{eq:D.38}
\end{align}
where
\begin{itemize}
\item[(g)] follows because $\mathcal{B}(M_{n}) \subseteq \tilde{\mathcal{A}}(M_{n})$,
\item[(h)] is due to \eqref{eq:4.28}.
\end{itemize}
\end{itemize}
From \eqref{eq:D.33} and \eqref{eq:D.38}, the variational distance between $\tilde{P}_{n}$ and $\tilde{Q}_{n}$ is bounded from above as
\begin{align}
d_{\mathrm{v}}(\tilde{P}_{n}, \tilde{Q}_{n}) 
&\le \tau + \tau \nonumber\\
&= 2\tau.
\end{align}
As a result, from Lemma \ref{lem:1} and the relation of each entropy,
\begin{align}
&|H(X_{\mathcal{K}}^{n} \in \mathcal{B}(J_{n})) - H(X_{\mathcal{K}}^{n} \in \tilde{\mathcal{A}}(J_{n}))|
\le -2\tau\log\frac{2\tau}{M_{n}}, \label{eq:D.41} \\
&|H(X_{\mathcal{H}}^{n}, X_{\mathcal{K}}^{n} \in \mathcal{B}(J_{n})) - H(X_{\mathcal{H}}^{n}, X_{\mathcal{K}}^{n} \in \tilde{\mathcal{A}}(J_{n}))| \nonumber\\
&\le  -2\tau\log\frac{2\tau}{|\mathcal{X}_{\mathcal{H}}|^{n} \cdot M_{n}}. \label{eq:D.42}
\end{align}
From \eqref{eq:D.41}, \eqref{eq:D.42}, and the chain rule of entropy,
\begin{align}
&|H(X_{\mathcal{H}}^{n} | X_{\mathcal{K}}^{n} \in \mathcal{B}(J_{n}){)} - H(X_{\mathcal{H}}^{n} | X_{\mathcal{K}}^{n} \in \tilde{\mathcal{A}}(J_{n}))|\nonumber\\
&= |\{H(X_{\mathcal{H}}^{n}, X_{\mathcal{K}}^{n} \in \mathcal{B}(J_{n})) - H(X_{\mathcal{K}}^{n} \in \mathcal{B}(J_{n}))\}\nonumber\\
&\hspace{4.5mm} - \{H(X_{\mathcal{H}}^{n}, X_{\mathcal{K}}^{n} \in \tilde{\mathcal{A}}(J_{n})) - H(X_{\mathcal{K}}^{n} \in \tilde{\mathcal{A}}(J_{n}))\}| \nonumber\\
&=  |\{H(X_{\mathcal{H}}^{n}, X_{\mathcal{K}}^{n} \in \mathcal{B}(J_{n}){)} - H(X_{\mathcal{H}}^{n}, X_{\mathcal{K}}^{n} \in \tilde{\mathcal{A}}(J_{n}))\}\nonumber\\
&\hspace{4.5mm} + \{H(X_{\mathcal{K}}^{n} \in \tilde{\mathcal{A}}(J_{n})) - H(X_{\mathcal{K}}^{n} \in \mathcal{B}(J_{n}))\}| \nonumber\\
&\stackrel{\rm{(i)}}{\le} |H(X_{\mathcal{H}}^{n}, X_{\mathcal{K}}^{n} \in \mathcal{B}(J_{n}){)} - H(X_{\mathcal{H}}^{n}, X_{\mathcal{K}}^{n} \in \tilde{\mathcal{A}}(J_{n}))|\nonumber\\
&\hspace{4.5mm} + |H(X_{\mathcal{K}}^{n} \in \tilde{\mathcal{A}}(J_{n})) - H(X_{\mathcal{K}}^{n} \in \mathcal{B}(J_{n}))| \nonumber\\
&\le  -2\tau\log\frac{2\tau}{M_{n}} - 2\tau\log\frac{2\tau}{|\mathcal{X}_{\mathcal{H}}|^{n} \cdot M_{n}} \nonumber\\
&=  -4\tau\log\frac{2\tau}{|\mathcal{X}_{\mathcal{H}}|^{n} \cdot M_{n}} \nonumber\\
&= 4\tau\log\frac{|\mathcal{X}_{\mathcal{H}}|^{n} \cdot M_{n}}{2\tau},
\end{align}
where
\begin{itemize}
\item[(i)] is because of the triangle inequality.
\end{itemize}
Therefore, we obtain
\begin{align}
\frac{1}{n}H(X_{\mathcal{H}}^{n} | J_{n}) 
&= \frac{1}{n}H(X_{\mathcal{H}}^{n} | X_{\mathcal{K}}^{n} \in \mathcal{B}(J_{n}))\nonumber\\
&\ge \frac{1}{n}H(X_{\mathcal{H}}^{n} | X_{\mathcal{K}}^{n} \in \tilde{\mathcal{A}}(J_{n})) - \frac{4\tau}{n}\log\frac{|\mathcal{X}_{\mathcal{H}}|^{n} \cdot M_{n}}{2\tau} \nonumber\\
&\stackrel{\rm{(j)}}{>} \frac{1}{n}H(X_{\mathcal{H}}^{n} | X_{\mathcal{K}}^{n} \in \tilde{\mathcal{A}}(J_{n})) - \frac{4\tau}{n}\log\frac{|\mathcal{X}_{\mathcal{H}}|^{n} \cdot 2^{nR}}{(2\tau)^{n}} \nonumber\\
&= \frac{1}{n}H(X_{\mathcal{H}}^{n} | X_{\mathcal{K}}^{n} \in \tilde{\mathcal{A}}(J_{n})) - 4\tau \log\frac{|\mathcal{X}_{\mathcal{H}}| \cdot 2^{R}}{2\tau},
\end{align}
where
\begin{itemize}
\item[(j)] follows from the definition that $M_{n} = 2^{nR}$ and $2\tau < 1$.
\end{itemize}
We complete the derivation of \eqref{eq:3.62}.\hspace{\fill}$\Box$

\section{Proof of \eqref{eq:3.68}}
First of all, we shall show
\begin{align}
x_{\mathcal{K}}^{n} \in \tilde{\mathcal{A}}(j) \Longrightarrow x_{\mathcal{R}}^{n} \in T_{\delta_{3}}^{n}(X_{\mathcal{R}} | x_{\mathcal{H}}^{n}, \hat{x}_{\mathcal{R}}^{n}(j)),\nonumber\\
\delta_{3} \coloneqq (|\mathcal{X}_{\mathcal{H}}| + 1) \cdot 2\delta. \label{eq:E.1}
\end{align}
By the definition of $\tilde{\mathcal{A}}(j)$,
\begin{align}
\tilde{\mathcal{A}}(j) \subseteq T_{2\delta}^{n}(X_{\mathcal{K}} | \hat{x}_{\mathcal{R}}^{n}(j))~~~\mathrm{for}~j = 1, 2, \ldots, M_{n}-1.
\end{align}
Thus, from Lemma \ref{lem:3}, any $x_{\mathcal{R}}^{n}$ such that $(x_{\mathcal{R}}^{n}, x_{\mathcal{H}}^{n}) \in \tilde{\mathcal{A}}(j)$ satisfies
\begin{align}
x_{\mathcal{R}}^{n} \in T_{\delta_{3}}^{n}(X_{\mathcal{R}} | x_{\mathcal{H}}^{n}, \hat{x}_{\mathcal{R}}^{n}(j)).
\end{align}
That is, given $x_{\mathcal{H}}^{n} \in \mathcal{X}_{\mathcal{H}}^{n}$ and $\hat{x}_{\mathcal{R}}^{n}(j) \in \hat{\mathcal{X}}_{\mathcal{R}}^{n}$, $x_{\mathcal{R}}^{n} \in \mathcal{X}_{\mathcal{R}}^{n}$ and $x_{\mathcal{K}}^{n} = (x_{\mathcal{R}}^{n}, x_{\mathcal{H}}^{n}) \in \tilde{\mathcal{A}}(j)$ are conditional strongly typical sequences. Then, we obtain \eqref{eq:E.1}, and
\begin{align}
&\sum_{x_{\mathcal{R}}^{n}\!:~(x_{\mathcal{R}}^{n}, x_{\mathcal{H}}^{n})\in\tilde{\mathcal{A}}(j)}\Pr\{X_{\mathcal{R}}^{n}=x_{\mathcal{R}}^{n}|X_{\mathcal{H}}^{n}=x_{\mathcal{H}}^{n}\}\Pr\{X_{\mathcal{H}}^{n}=x_{\mathcal{H}}^{n}\}\nonumber\\
\le &\sum_{x_{\mathcal{R}}^{n}\in T_{\delta_{3}}^{n}(X_{\mathcal{R}}|x_{\mathcal{H}}^{n}, \hat{x}_{\mathcal{R}}^{n}(j))}\Pr\{X_{\mathcal{R}}^{n}=x_{\mathcal{R}}^{n}|X_{\mathcal{H}}^{n}=x_{\mathcal{H}}^{n}\}\Pr\{X_{\mathcal{H}}^{n}=x_{\mathcal{H}}^{n}\}.
\end{align}
Therefore, we obtain \eqref{eq:3.68}.\hspace{\fill}$\Box$

\end{document}